\documentclass[11pt,a4paper]{article}
\usepackage{cite}
\usepackage{a4}

\usepackage{graphicx}
\usepackage{amsmath}
\usepackage{amssymb}

\begin{document}

\title{Two band superconductivity in MgB$_2$: basic anisotropic
properties and phase diagram}
\author{Manuel Angst$^{1,2}$ and Roman Puzniak$^3$\\
$^{1}$ Physik-Institut der Universit{\"a}t Z\"{u}rich, 8057
Z\"urich, Switzerland \\
$^{2}$ Solid State Physics Laboratory,  ETH Z\"{u}rich, 8093 Z\"{u}rich, Switzerland\\
$^{3}$ Institute of Physics, Polish Academy of Sciences,\\ Al.
Lotnikow 32/46, 02-668 Warsaw, Poland }

\maketitle

\begin{abstract}
Magnesium diboride MgB$_2$ has been an extraordinarily ``hot''
research topic in the two years since the discovery of
superconductivity below $T_{\mathrm{c}}\approx 40\,{\mathrm{K}}$
in this compound in $2001$. A large part of the excitement is due
to the vast amount of unusual properties originating from the
involvement in superconductivity of two sets of bands that are of
a very different nature. Because of the different dimensionality
of the two sets of bands, this ``two band superconductivity''
leads two a complex behavior of the anisotropic superconducting
state properties as a function of magnetic field and temperature.
The interplay between two band superconductivity and the
anisotropic mixed state properties and phase diagram is reviewed,
with a strong focus on experimental results obtained by torque
magnetometry on MgB$_2$ single crystals. The different
dimensionality of the two sets of bands is manifested in the upper
critical field $H_{\mathrm{c2}}$ anisotropy $\gamma_H$, which was
found to strongly decrease with increasing temperature. While the
angular dependence of $H_{\mathrm{c2}}$ follows roughly the
predictions of anisotropic Ginzburg-Landau theory AGLT, small, but
systematic deviations were observed near $T_{\mathrm{c}}$. This,
and the temperature dependent $\gamma_H$, witness a breakdown of
the AGLT description of MgB$_2$, even close to $T_{\mathrm{c}}$.
Theoretical calculations are in qualitative agreement with the
observed $\gamma_H (T)$ dependence, but suggest a difference
between $\gamma_H$ and the low field penetration depth anisotropy
$\gamma_\lambda$ with much lower values of $\gamma_\lambda$ at low
temperatures. Measurements of reversible torque vs angle curves in
the mixed state on the one hand are incompatible with expected
``single anisotropy behavior''. On the other hand, they indicate
that the penetration depth anisotropy $\gamma_\lambda$ in
intermediate fields has to be higher than currently available
theoretical estimates. The field dependent effective anisotropies
observed in torque as well as many other experiments can be
accounted for qualitatively by the faster depression of
superconductivity in the more isotropic bands by the applied
field. The vortex matter phase diagram is drawn, based on
measurements of the reversible and irreversible torque in the
mixed state. In the irreversible torque, a peak effect (PE) was
observed in fields of about $0.85\,H_{\mathrm{c2}}$. History
effects of the critical current density in the PE region suggest
that the PE signifies an order-disorder transition of vortex
matter, analogous to findings in low as well as high
$T_{\mathrm{c}}$ superconductors.
\end{abstract}

\newpage

\section{Magnesium Diboride: an unconventional ``conventional superconductor''}

\begin{figure}[!bt]
\centering
\includegraphics[width=0.45\linewidth]{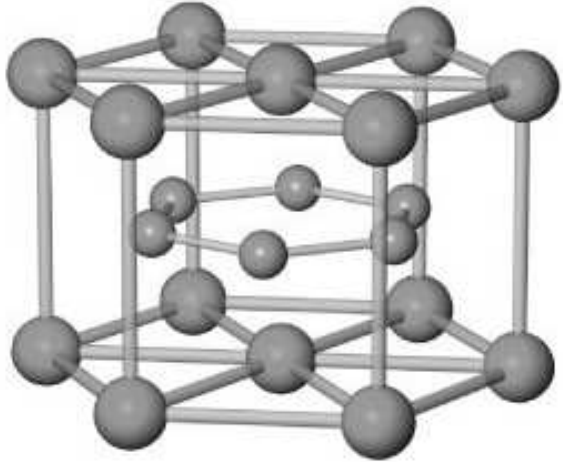}
\caption{Crystal structure of magnesium diboride MgB$_2$. The
structure consists of alternating layers of magnesium (large
spheres) and boron (small spheres) atoms. Figure from Ref.\
\cite{Larbalestier01rev}, $\copyright$ copyright Nature Publishing
group.} \label{struc}
\end{figure}

Magnesium diboride MgB$_2$ is a binary inter-metallic compound
\cite{Russell53} with a simple, layered structure: Boron atoms
form a honeycomb lattice with hexagonal magnesium layers
sandwiched between the B layers (see Fig.\ \ref{struc}). The
material has been known since the 1950's \cite{Russell53}, but
little attention was paid to this compound. This changed
dramatically in January 2001, when it was reported that MgB$_2$
becomes superconducting below $T_{\text{c}}\approx 39 \,
{\text{K}}$ \cite{Nagamatsu01}. Particularly the high transition
temperature caused a lot of interest in this compound and it's
physical properties. Just ten months after the discovery of
superconductivity in MgB$_2$, a comprehensive review
\cite{Buzea01} appeared, covering the first months of research
about this ``new'' superconductor. The review counted 263 studies
that between January and July either appeared in journals or were
available to the community as e-prints.

The initial excitement was driven by the high transition
temperature, thought to be close to the limit of what can be
expected for a material exhibiting superconductivity with a
conventional, phonon mediated pairing mechanism
\cite{note_phononTc}. The high $T_{\text{c}}$ of MgB$_2$ therefore
immediately raised speculations about an unconventional pairing
mechanism. However, studies of the boron isotope effect on
$T_{\mathrm{c}}$ \cite{Budko01} and, e.g., the $^{11}$B nuclear
spin-lattice relaxation rate \cite{Kotegawa01} soon indicated a
BCS type s-wave phonon-mediated mechanism of superconductivity.
The rather conventional nature of the superconducting pairing in
MgB$_2$ is also evidenced by the absence of a boron isotope effect
on the penetration depth \cite{DiCastro03}, in contrast to
observations on cuprate superconductors \cite{note_Keller03}.

On the other, MgB$_2$ is far from being just an ``all ordinary''
conventional superconductor well described by BCS theory, as many
other experiments showed. Point contact spectroscopy
\cite{Schmidt01} indicated, for example, an energy gap to
$T_{\text{c}}$ ratio much lower than predicted by BCS theory, the
results of specific heat measurements \cite{Wang01c,Bouquet01b}
were also at odds with BCS predictions, and the total (Mg and B)
isotope effect on $T_{\mathrm{c}}$ was found to be substantially
lower than expected as well \cite{Hinks01}.

The huge difference between the effects of boron and magnesium
isotope substitution \cite{Hinks01} provided an early clue for a
highly selective electron-phonon coupling (EPC), and the specific
heat measurements \cite{Wang01c,Bouquet01b} even indicated the
existence of a second superconducting gap. Calculations of the
band-structure and electron-phonon interaction led to the
description of MgB$_2$ as a peculiar ``two band'' superconductor
\cite{Liu01,Choi02,Choi02b}, whose features show up in almost
every conceivable measurement \cite{Canfield03}.

The chapter is organized as follows. In Sec.\ \ref{twoband}, the
concept and origin of two band superconductivity in MgB$_2$ are
sketched, as well as its experimental confirmation. One of the
features heavily influenced by the two band nature of
superconductivity is the angular and temperature dependence of the
upper critical field $H_{\mathrm{c2}}$, related to the coherence
length. In Sec.\ \ref{Hc2}, torque measurements of
$H_{\mathrm{c2}}(T,\theta)$ are presented and discussed. Section
\ref{Hc2} also contains a brief discussion of anisotropic
Ginzburg-Landau theory and basic experimental procedures (see
Angst \cite{AngstPhD} for a more comprehensive discussion). The
question of how exactly two band superconductivity shapes the
anisotropy of the basic length scales below the (bulk)
$H_{\mathrm{c2}}$, is still not completely solved. In Sec.\
\ref{mixedstate}, we review torque (and selected other)
measurements in the mixed state and compare to recent theoretical
concepts and predictions. Apart from two band superconductivity,
MgB$_2$ is also interesting in that thermal fluctuations of
vortices in this superconductor are of a strength intermediate
between the ones of the cuprate and of conventional
superconductors. In Sec.\ \ref{PE}, we present and discuss the
vortex matter phase diagram, drawn based on torque measurements.
Finally, in Sec.\ \ref{conc}, we summarize our discussion and
present the main conclusions.

\section{Two band superconductivity: a text book example}
\label{twoband}

A two band scenario leading to two superconducting gaps was
already proposed just two years after formulation of the BCS
theory \cite{Suhl59}. Two different bands in which
superconductivity can exist are the most essential ingredient.
Without any scattering of electrons by phonons from one band to
the other one, there would be two transition temperatures. In the
case of weak, but finite, interband phonon scattering, the lower
$T_{\text{c}}$ disappears and the temperature dependence of the
lower gap becomes strongly non-BCS. With very large phonon
interband scattering, there should be still two gaps with
temperature dependences now roughly similar to BCS predictions.
The initial two-gap proposal was made in view of applications to
$s-d$ metals. However, while some hints of ``two-gap effects''
were experimentally thought to be detected in form of small
deviations from BCS in transitions metals \cite{Carlson70}, no
strong effects, unequivocally attributable to two superconducting
gaps, were found. It was pointed out \cite{Butler76} that, in
order to observe considerable two-band effects, it was not enough
to have two overlapping bands crossing the Fermi level. An
additional requirement was that the bands in question have very
different physical origins, such as covalent vs metallic-type
bonding, which is more likely to occur in compounds.

The most convincing historical demonstration of two-gap
superconductivity was made by tunnelling experiments in Nb doped
SrTiO$_3$ \cite{Binnig80}, where clear double peaks in the
tunnelling conductance were found. Two- or multi- band models were
also proposed for the cuprate H$T_{\text{c}}$SC (see, e.g., Kresin
{\em et al.} \cite{Kresin90}), but strong electron-electron
correlations and magnetism make those materials quite difficult to
tract with first principles calculations \cite{Pickett89}, and
unequivocal experimental evidence for any observable effects being
due to multi band effects is scarce. It was only after the
discovery of superconductivity in magnesium diboride that strong
two-band effects were seen by a large number of different
experimental techniques. What is it that really makes MgB$_2$ so
special?

Answering this question was helped tremendously by the fact that
the relatively simple crystal structure (see Fig.\ \ref{struc})
and the lack of complications arising from spin-dependent
interactions or strong electron-electron correlations made
accurate {\em ab initio} calculations of the electronic structure
\cite{Belashchenko01,An01,Kortus01} and of the lattice dynamics
\cite{Bohnen01,Kong01,Yildirim01} possible. In the remainder of
the section we briefly sketch this answer, about which by now a
general consensus developed.

Magnesium atoms in MgB$_2$ are effectively ionized
\cite{Belashchenko01}, i.e., they serve mainly as electron donors,
similar to electron doping of graphite by intercalation. Like
carbon atoms in graphite, boron atoms are sp$^2$ hybridized, with
three of the four \cite{note_gainedel} valence electrons tied up
in strong covalent $\sigma$ ($p_{x,y}$) bonds lying in plane,
while the fourth electron is in nonbonding $\pi$ ($p_z$) states,
which are delocalized (like the $\pi$ bonds in benzene).

Apart from doping the boron sublattice with electrons, Mg atoms,
respectively Mg$^{2+}$ ions, have another crucial impact on the
electronic structure of MgB$_2$ : Since the Mg$^{2+}$ attractive
potential is felt more strongly by the electrons in the $\pi$
states than by the ones in the $\sigma$ states, magnesium induces
charge transfer from the $\sigma$ to the $\pi$ states, accompanied
by the corresponding relative shift of the $\sigma$ and $\pi$ band
energies.

\begin{figure}[tb]
\centering
\includegraphics[width=0.6\linewidth]{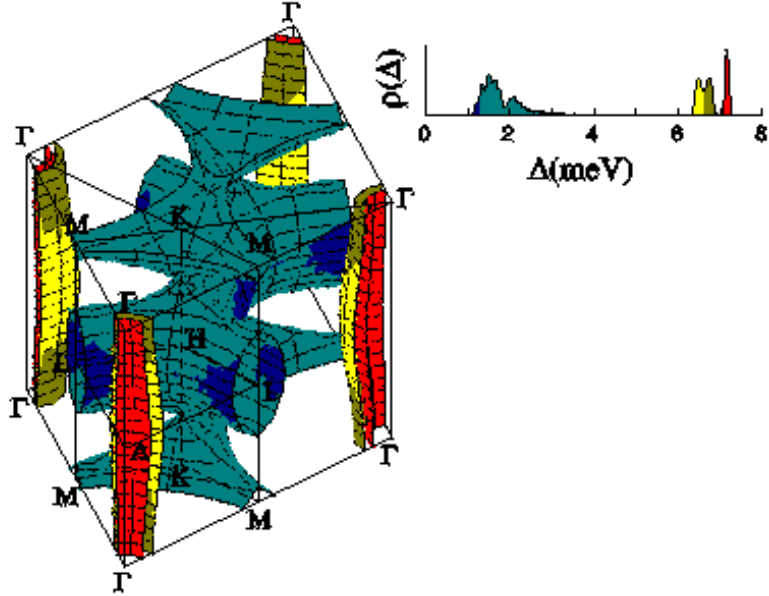}
\caption{ Fermi surface of MgB$_2$, consisting of four separate
sheets: the tubular networks in the center of the figure are the
two $\pi$ sheets, the two $\Gamma$ centered (approximate)
cylinders the $\sigma$ sheets (see text). The coloring/shading
indicates the wave vector dependent energy gap of MgB$_2$ without
any impurity scattering at $4\,{\mathrm{K}}$, with the coding and
the distribution of gap values given besides the main figure. On
the nearly-cylindrical $\sigma$ sheets, the gap value is $\sim
6.5-7.5\,{\mathrm{meV}}$, while on the 3D $\pi$ sheets it is $\sim
1-3\,{\mathrm{meV}}$. Figure from Ref.\ \cite{Choi02b},
$\copyright$ copyright Nature Publishing group.} \label{FS}
\end{figure}

This leads the $\sigma$ bands of MgB$_2$ to be incompletely
filled, in contrast to graphite. The covalent $\sigma$ bonds are
``driven metallic'', i.e. the corresponding carriers become
mobile, by Mg induced charge transfer. The calculated Fermi
surface (FS), shown in Fig.\ \ref{FS}, consists of four separate
sheets. Two sheets, derived from the boron $\pi$ ($p_z$) bands,
are (electronlike and holelike) tubular networks. Since the
smaller B$-$B in plane distance is compensated by a smaller
($pp\pi$ vs $pp\sigma$) hopping \cite{Kortus01}, they are
virtually isotropic and clearly three-dimensional (3D). The other
two sheets, derived from boron $\sigma$ ($p_{x,y}$) bands are
holelike (approximate) cylinders with axis $\Gamma \! - \!
\mathrm{A}$. The $\sigma$ sheets show very little dispersion along
$\Gamma \! - \! \mathrm{A}$ and are thus nearly two-dimensional
(2D), which could be expected from the covalent directed nature of
the associated intraplanar B$-$B bonds. Correspondingly, a high
anisotropy of the Fermi velocity, averaged over the $\sigma$
sheets, was calculated \cite{Belashchenko01}: $\langle v_{ab}^2
\rangle _{\sigma} / \langle v_{c}^2 \rangle _{\sigma} \simeq 46$,
where $\langle \ldots \rangle _{\sigma}$ denotes the average over
the $\sigma$ Fermi sheets. The quasi two-dimensionality of the
$\sigma$ bands also leads to a sizeable contribution to the
density of states (DOS) at the Fermi level $-$ despite the very
small hole doping level \cite{An01}. In fact, the contributions of
the 2D $\sigma$ bands to the total DOS at the FS is almost equal
in size to the one of the 3D $\pi$ bands (44\% vs 56\%
\cite{Choi02b}). This feature of two sets of bands with different
dimensionality, but comparable contribution to the DOS at the
Fermi level, is very peculiar.

For phonon mediated superconductivity, the phonon ``band
structure'' and the EPC are just as important as the details of
the structure of the electronic subsystem. The layered structure
of MgB$_2$ is reflected in the phonon dispersion curves as well,
by weak dispersion of the optical branches along $\Gamma -
\mathrm{A}$, and by anisotropic acoustic branches. Of the four
distinct branches with non zero energy at the zone center
$\Gamma$, three were found to have no sizeable coupling to the
electronic subsystem. This is not the case, however, for the
doubly degenerate mode with $E_{2g}$ symmetry at $\Gamma$, which
involves only in-plane (hexagon distorting) movement of boron
atoms and is strongly anharmonic \cite{Yildirim01,Boeri02}: This
movement leads to large variations in the overlap of orbitals of
neighboring B atoms, obviously modulating bond energies,
particularly of the covalent $\sigma$ bonds. Detailed zone-center
frozen phonon calculations \cite{Yildirim01} indeed find a
coupling of the $E_{2g}$ phonon mode to the $\sigma$ bands, which
is not only very large, but also strongly non-linear, suggesting
even pairing via two-phonon exchange \cite{Liu01}. This
anomalously large coupling is evidenced, for example, by an
extremely large linewidth of the $E_{2g}$ mode \cite{Shukla03}. It
is important to note that the large and nonlinear coupling, the
anharmonicity, as well as the small carrier density in the
$\sigma$ bands \cite{note_nonadiab} are intimately related and
have competing effects on $T_{\mathrm{c}}$ \cite{Mazin03}.

With such an anisotropic (sheet$-$ or, more general,
$\overrightarrow{k} \! -$dependent) EPC, the superconducting
pairing (and order parameter, gap) should be anisotropic as well,
at least in the clean limit. An analysis of pairing, decomposed
into the four Fermi sheet contributions, led Liu {\em et al.}
\cite{Liu01} to propose the two-gap scenario for MgB$_2$ : Similar
anisotropic Fermi velocities and EPC of the two $\sigma$ Fermi
sheets, and of the two $\pi$ Fermi sheets, respectively, allow the
reduction to an effective two-band model. Allowing for different
order parameters for each of these two ($\sigma$ and $\pi$)
effective bands increases the effective coupling constant relevant
for superconductivity and is, together with the extremely strong
coupling between one phonon mode and the $\sigma$ bands,
responsible for the high $T_{\mathrm{c}}$ of MgB$_2$
\cite{note_anisTc}. Later numerical calculations using fully
anisotropic Eliashberg theory \cite{Choi02b,Choi02} qualitatively
confirmed the simpler effective two band model by Liu {\em et
al.}. The gap distribution of gap values over the Fermi surface
resulting from these calculations is shown in Fig.\ \ref{FS}. Note
that these results apply for hypothetical MgB$_2$ without any
impurity scattering \cite{Mazin03}. In the 2D $\sigma$ band, the
superconducting gap-to-$T_{\text{c}}$ ratio is slightly enhanced
with respect to BCS theory, while the gap in the 3D $\pi$ band is
about two to three times smaller, and the corresponding
gap-to-$T_{\text{c}}$ ratio is substantially below BCS
predictions.

In order to have two superconducting gaps staying open up to the
bulk $T_{\mathrm{c}}$, it is not only necessary to have a sizeable
coupling between the two sets of bands by scattering of carriers
by phonons. In addition, scattering of carriers between the two
sets of bands by impurities needs to be very low; otherwise, any
gap differences would be ``washed out'' by impurity scattering,
accompanied by a $T_{\mathrm{c}}$ lowering \cite{Golubov97}. In
MgB$_2$, the $\sigma-\pi$ impurity scattering is fortunately very
small even in rather dirty samples because $\sigma$ and $\pi$
bands are formed from different local orbitals {\em orthogonal on
an atomic scale} and therefore $\sigma-\pi$ hybridization is very
small \cite{Mazin02}. On the other hand, the {\em intraband}
variation of the gaps as calculated for the clean case
\cite{Choi02b} and shown in Fig.\ \ref{FS} was argued to be
completely washed out in any realistic sample due to intraband
impurity scattering \cite{Mazin03}.

The two band two gap scenario is meanwhile well verified
experimentally, from a large number of measurements both on
polycrystalline material and single crystals: The band structure
calculations are in agreement with de~Haas-van~Alphen
\cite{Yelland02,Carrington03} and angle resolved photoemission
spectroscopy \cite{Uchiyama02} results, and the strong coupling
between the $E_{2g}$ phonon mode and the $\sigma$ bands is
evidenced by inelastic neutron scattering \cite{Yildirim01} and
Raman \cite{Bohnen01} measurements: Spectroscopic measurements
with various techniques, including point contact spectroscopy
\cite{Szabo01,Laube01,Gonnelli02}, scanning tunnelling
spectroscopy \cite{Giubileo01,Giubileo01a,Iavarone02,Eskildsen02},
photoemission spectroscopy \cite{Tsuda01}, break junction
(``S$-$I$-$S'') tunnelling spectroscopy \cite{Schmidt02}, and
Raman scattering \cite{Chen01,Quilty02} observe directly two gaps.
Moreover, the two different gaps and their location in $k$ space
on the $\sigma$ and $\pi$ Fermi sheets have been directly observed
recently by angle-resolved photoemission spectroscopy (ARPES)
\cite{Souma03}. The unusual temperature dependence of (zero field)
thermodynamic properties of the superconducting state, as seen by
bulk probes, particularly specific heat
\cite{Wang01c,Bouquet01b,Bouquet01c} and thermal conductivity
\cite{Sologubenko02b}, can also be accounted for quantitatively
\cite{Choi02b,Bouquet01c,Golubov02} with two-gap
superconductivity, as can the incomplete isotope effect
\cite{Budko01,Choi02,Hinks01}. Due to the existence of two,
coupled, superconducting condensates with different energy and
length scales as well as very different anisotropies, a variety of
interesting and unusual effects can be expected in the vortex
state in non zero magnetic field, as will be discussed in the next
sections, which focus on results by torque magnetometry.

In summary, magnesium diboride is peculiar in that two sets of
bands with {\em very different average gap values} contribute to
the DOS at the FS {\em with similar strengths}, and that they have
{\em different ``dimensionalities''} (or more accurately very
different Fermi velocity anisotropies) and {\em different coupling
strengths to the phonon subsystem}. For the actual observation of
two band superconductivity, it is also crucial that there is some
coupling, but at the same time extremely low impurity scattering
between the two sets of bands. Two band superconductivity in
MgB$_2$ leads to very peculiar effects showing up in a vast range
of experiments. Experimental results in zero field can be well
explained quantitatively within the two gap model.

\section{Upper critical field $H_{\mathrm{c2}}$: breakdown of anisotropic Ginzburg-Landau theory}
\label{Hc2}

In MgB$_2$, as we have seen in Sec.\ \ref{twoband}, the crystal
structure as well as the electronic and phononic band structure
are all far from isotropic. This should lead to anisotropic
superconducting state properties as well, for example to an
anisotropic upper critical field $H_{\mathrm{c2}}$. Early
measurements on polycrystalline or thin film MgB$_2$
\cite{Simon01,Budko01b,Delima01,Lima01a,Papavassiliou02,Buzea01}
as well as electrical transport measurements on single crystals
\cite{Kim02,Xu01,Lee01,Pradhan01} indeed found such anisotropy,
but with a very wide span of values reported.

The upper critical field is the maximum field for which
superconductivity in the mixed state can persist. The mixed state
becomes energetically unfavorable when the vortex cores overlap
too strongly, since the order parameter is suppressed in the
cores. Considering the isotropic case, the vortex core radius is
given by the coherence length $\xi$, the typical length scale for
variations of the order parameter, which (ignoring the difference
between GL and BCS coherence lengths) at zero temperature is
related to the superconducting gap by
\begin{equation}
\xi_{\circ}=\hbar v_{\mathrm{F}}/\pi \Delta (0), \label{xigap}
\end{equation}
where $v_{\mathrm{F}}$ is an average velocity at the Fermi
surface. The upper critical field is given by
\begin{equation}
H_{\mathrm{c2}}=\Phi_{\circ}/2\pi\xi^2 \label{xiHc2}
\end{equation}
with the flux quantum $\Phi_{\circ}=2.07\times
10^{-7}\,{\mathrm{Gcm}}^2$ \cite{Tinkham_intro}.

Anisotropic superconductors are usually treated within the
phenomenological {\em anisotropic Ginzburg-Landau theory} (AGLT),
which is obtained from the isotropic Ginzburg-Landau (GL) theory
through replacing the effective mass \cite{note_effmass} $m^*$ in
the GL free energy functional
\cite{Ginzburg_Landau50,Tinkham_intro} by an effective mass {\em
tensor}, with values $m_a^*$, $m_b^*$, and $m_c^*$ along the
principal axes \cite{Ginzburg52,Caroli63,Gorkov64}.
%Mostly,
%these effective masses are derived from measurements of
%superconducting state properties, but they are usually thought to
%be equal to the band effective masses \cite{note_effmass}.
In the most usual case of uniaxial anisotropy, all anisotropy is
incorporated into the anisotropy parameter
$\gamma=(m_c^*/m_{ab}^*)^{1/2}$, a simple constant. From the free
energy variation it follows that
\begin{equation}
\gamma=(m_c^*/m_{ab}^*)^{1/2}=\lambda_{c}/\lambda_{ab}=\xi_{ab}/\xi_{c}=H_{\text{c2}}^{\|ab}/H_{\text{c2}}^{\|c}\equiv
\gamma_H. \label{gammas}
\end{equation}
Here, $\lambda$ is the penetration depth, the typical length scale
for variations of the (local) magnetic field, given by
\begin{equation}
\lambda ^{-2} = \frac{4 \pi e^2}{c^2} \frac{n_s}{m^{*}},
\label{lambdadef}
\end{equation}
where $n_s$ is the superfluid density (density of superconducting
carriers), $m^{*}$ is the (band averaged) effective mass tensor
\cite{note_effmass}, $e$ is the electron charge and $c$ the speed
of light.

When the field $H$ is not applied parallel or perpendicular to the
$c$-axis or the $ab$-plane, scaling relations apply, for example
the AGLT angular dependence of the upper critical field is given
by
\begin{equation}
H_{\text{c2}}^{\mathrm{AGL}}(\theta) = H_{\text{c2}}^{\|c}/ (
\cos^2 \theta + \sin^2 \theta / \gamma_H^2 )^{1/2} \equiv
H_{\text{c2}}^{\|c}/\epsilon (\theta), \label{Hc2_theta}
\end{equation}
where $\theta$ is the angle between the field and the $c$-axis of
the sample \cite{Tilley65,Blatter92}. Since effective masses
\cite{note_effmass} are temperature and field independent, at
least as long as polaronic effects and the thermal expansion of
the lattice can be neglected, the anisotropy parameter $\gamma$ is
assumed to be constant as well.

In materials with complex Fermi surfaces, the situation may not be
as simple as outlined above. For example, slight temperature
dependences of the $H_{\mathrm{c2}}$ anisotropy, which we call
$\gamma_H$ hereafter, were observed in some materials, such as
NbSe$_2$ \cite{Muto77}, and explained by microscopic theories
taking into account non-local effects
\cite{Takanaka75,Teichler75}. These theories show that the
temperature dependence of $\gamma$ cannot be attributed to an
anisotropy of the band effective mass tensor, unless it is also
wave-vector dependent. An anisotropic energy gap, caused by an
anisotropic attractive electron-electron interaction, however, can
lead to variations of $\gamma$ with $T$.

In MgB$_2$, the situation is particularly complex, due to the two
sets of bands: the $\pi$ band has a nearly isotropic Fermi
velocity and a small superconducting gap, which would lead to an
isotropic coherence length of the order of $\xi_\circ
^{\pi}\approx 50\,{\mathrm{nm}}$. The highly anisotropic $\sigma$
band with a larger gap would have a coherence length anisotropy of
about 6, with $\xi_\circ ^{\sigma,ab}\approx 10\,{\mathrm{nm}}$.
Which set of bands controls the bulk upper critical field and it's
anisotropy $\gamma_H$?

\begin{figure}[tb]
\centering
\includegraphics[width=0.85\linewidth]{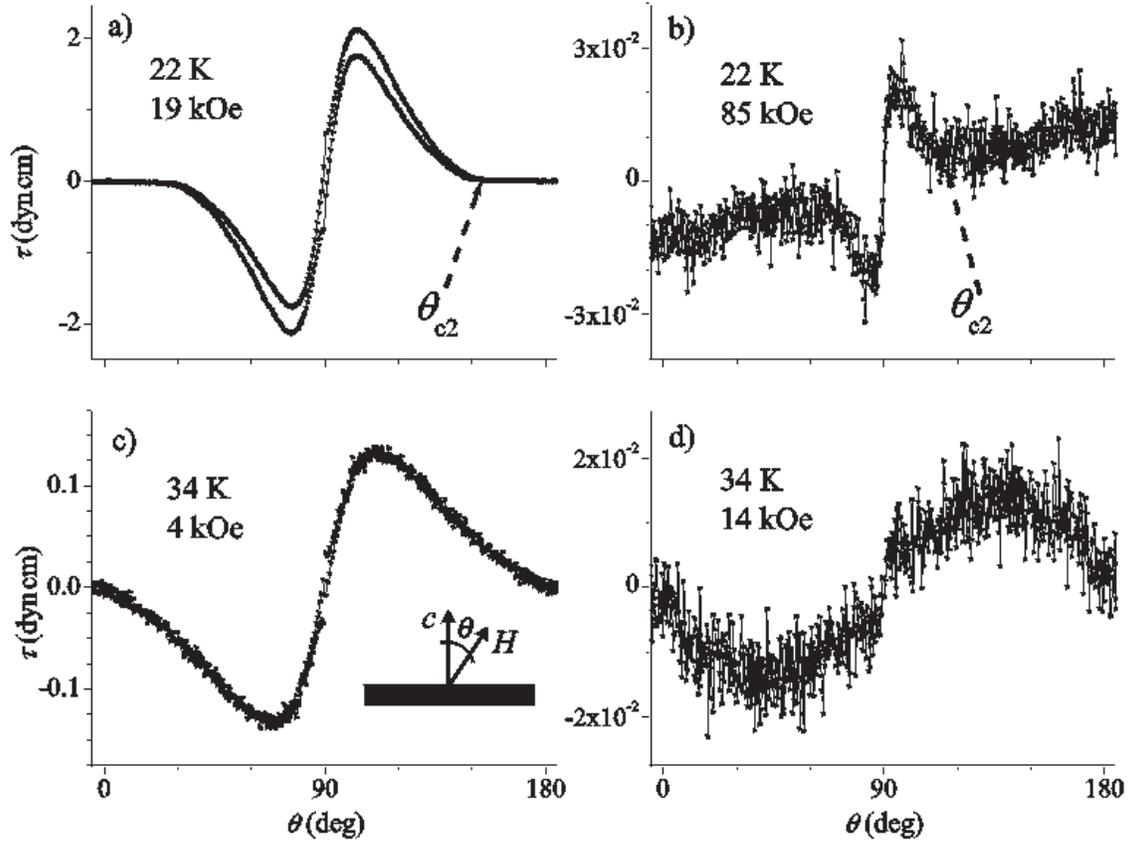}
\caption{ Torque $\tau$ vs.\ angle $\theta$ data of an MgB$_2$
single crystal (B), under four different conditions. The raw data
have been antisymmetrized around $90\,{\mathrm{deg}}$ in order to
subtract a symmetric background. $\theta_{\mathrm{c2}}$ indicates
the angle for which the applied field is the upper critical field.
The schematic drawing in c) shows the definition of the angle
$\theta$. After Ref.\ \cite{Angst03MgB2PhyC}.} \label{rawcurves}
\end{figure}

The torque magnetometry results reviewed in this section
\cite{MgB2anisPRL02,Karpinski03SST,Angst03MgB2PhyC,Angst03Rio}, as
well as several other experimental
\cite{Sologubenko02,Budko02,Eltsev02,Ferdeghini02,Zehetmayer02,Lyard02,Welp03,Machida03}
and theoretical
\cite{Posazhennikova02,Miranovic03,Gurevich03,Dahm02,Golubov03}
studies indicate that the answer strongly depends on temperature:
the $T$ dependence of $\gamma_H$ in MgB$_2$ is much more
pronounced than ever observed before in any compound.

In order to precisely determine the angle dependence and the
anisotropy of the upper critical field of MgB$_2$, we carried out
torque measurements in fields up to $90\,{\mathrm{kOe}}$ on
several MgB$_2$ single crystals with volumes between about
$4\times 10^{-4}\, {\text{mm}}^3$ and $2\times 10^{-2}\,
{\text{mm}}^3$ \cite{note_ABC}. Crystals of MgB$_2$ were grown
with a high pressure cubic anvil technique, described in detail by
Karpinski {\em et al.} \cite{Karpinski03SST,Karpinski03phyc}. In
brief, a mixture of Mg and B was put into a BN container and a
pressure of 30$-$35$ \, {\text{kbar}}$ was applied. Growth runs
consisted of heating during $1 \, {\text{h}}$ up to the maximum
temperature of 1700$-$1800$^{\circ}{\text{C}}$, keeping the
temperature for 1$-$3$ \, {\text{h}}$ and then cooling to room
temperature during 1$-$2$ \, {\text{h}}$. Flat crystals were up to
$0.8\times 0.6\times 0.05 \, {\text{mm}}^3$ in size, with sharp
transitions at about 38$-$39$ \, {\text{K}}$. The crystals
selected for torque measurements had $T_{\text{c}}\simeq 38 \,
{\text{K}}$: See Fig.\ 1 in Ref.\ \cite{MgB2PE} for a low field
magnetization curve in the region of the superconducting
transition of one of the crystals used for the torque study
(crystal B \cite{note_ABC}).

The torque $\vec{\tau}=\vec{m}\times \vec{B}\simeq \vec{m}\times
\vec{H}$, where $\vec{m}$ is the magnetic moment of the sample,
was recorded as a function of the angle $\theta$ between applied
field and the $c-$axis of the crystal for various fixed
temperatures and fields or as a function of the magnetic field at
fixed angles and temperatures. Measurements were performed on
miniaturized piezoresistive cantilevers specifically designed for
torque magnetometry \cite{Willemin98b}. For measurements close to
$T_{\mathrm{c}}$ in fields up to $14 \, {\text{kOe}}$, a
non-commercial magnetometer with very high sensitivity was used.
For the measurements in the mixed state described in Sec.\
\ref{mixedstate}, a vortex-shaking process by an additional small
oscillating magnetic field perpendicular to the main field was
employed to speed up the relaxation of the vortex lattice into
it's reversible state \cite{Willemin98,Willemin98a}. The crystal
labelled A was measured in this system. The other two crystals
were measured in a wider range of temperatures down to $15\,
{\text{K}}$ in a Quantum Design PPMS with torque option and a
maximum field of $90\, {\text{kOe}}$.

Four examples of torque vs angle curves are given in Fig.\
\ref{rawcurves}. Panels a) and b) correspond to measurements at
$22\,{\text{K}}$. For fields nearly parallel to the $c$-axis, both
curves are flat, apart from a small background visible in panel
b). Only when $H$ is nearly parallel to the $ab-$plane there is an
appreciable torque signal. The curve can be interpreted in a
straight-forward way: for $H$ parallel to the $c-$axis the sample
is in the normal state, while for $H$ parallel to the $ab-$plane
it is in the superconducting state. The crossover angle
$\theta_{\text{c2}}$ between the normal and the superconducting
state is the angle for which the fixed applied field is the upper
critical field. From the existence of both superconducting and
normal angular regions follows immediately that
$H_{\text{c2}}^{\|c}(22\,{\text{K}})<19\,{\text{kOe}}$ and
$85\,{\text{kOe}}<H_{\text{c2}}^{\|ab}(22\,{\text{K}})$. In panel
c), on the other hand, the crystal is seen to be in the
superconducting state for all values of the angle $\theta$, and
therefore $4\,{\text{kOe}}<H_{\text{c2}}^{\|c}(34\,{\text{K}})$.
Finally, the data in panel d) show only a small background
contribution \cite{note_incompatible}. Therefore, the crystal is
here in the normal state for any $\theta$, and we have
$H_{\text{c2}}^{\|ab}(34\,{\text{K}})<14\,{\text{kOe}}$.

From figure \ref{rawcurves} we therefore have two limitations for
the upper critical field anisotropy $\gamma_H$, without any
detailed $H_{\text{c2}}$ criterion, and without any model fits :
\begin{equation}
\gamma_H (22\,{\text{K}})>\frac{85}{19}\simeq 4.47; \; \gamma_H
(34\,{\text{K}})<\frac{14}{4}= 3.5. \label{rawproof}
\end{equation}
With the limitations of Eq.\ (\ref{rawproof}), {\em the upper
critical field anisotropy $\gamma_H$ of MgB$_2$ cannot be
temperature independent}! As a consequence, the standard {\em
anisotropic Ginzburg-Landau theory} AGLT {\em does not apply for
MgB$_2$}. The deviation is rather pronounced, within a change of
temperature of about $0.3\,T_{\text{c}}$, $\gamma_H$ changes, {\em
at least}, by a fifth of it's value. Concerning the real values of
$\gamma_H$ and $H_{\text{c2}}(\theta)$, we only have an estimation
so far and a more detailed analysis is necessary.

Although, in the light of the discussion of Fig.\ \ref{rawcurves},
it is clear that AGLT, with it's effective mass anisotropy model,
is not able the describe the data measured at {\em different
temperatures} consistently, the detailed analysis of the angle
dependence of $H_{\text{c2}}$ is based on AGLT. We will show that
as long as we stay at a {\em fixed temperature} well below
$T_{\mathrm{c}}$, AGLT is able to describe $H_{\text{c2}}(\theta)$
remarkably well \cite{note_AHLTfixedT}. Although the location of
$\theta_{\text{c2}}$, for example in Fig.\ \ref{rawcurves}a),
seems clear at first sight, this clarity disappears, when
examining the transition region in a scale necessary for the
precise determination of $\theta_{\text{c2}}$ (see Fig.\ 1 in
Ref.\ \cite{MgB2anisPRL02}). For a strict analysis, it is
necessary to take into account that the transition at
$H_{\text{c2}}$ is rounded by fluctuations of the order parameter
around the value minimizing the free energy.

%We therefore make a quick excursion to the theory of
%superconducting fluctuations. The brief account of the
%phenomenological Ginzburg-Landau theory and AGLT given in Sec.\
%\ref{GL} followed a ``mean field'' treatment, i.e., the order
%parameter $\Psi(\overrightarrow{r})$ was taken as always being
%equal to the order parameter $\Psi_{\circ}(\overrightarrow{r})$
%with minimum free energy. However, thermal fluctuations allow the
%system to sample order parameters with non-minimal free energy $-$
%in fact, any function $\Psi(\overrightarrow{r})$ with an energy
%difference to the minimum free energy of less than about $kT$ has
%significant statistical weight in determining the superconducting
%properties. There are measurable deviations from mean field
%behavior even in low $T_{\text{c}}$ compounds, but the deviations
%are small outside of a narrow transition region (see, e.g., Ref.\
%\cite{Tinkham_intro}). This situation is, however, completely
%different for the H$T_{\text{c}}$SC cuprates, where the
%unambiguous experimental definition of $H_{\text{c2}}$ is
%difficult because of the large influence of fluctuations.
To assess the importance of thermal fluctuations in a
superconductor, the so-called Ginzburg number
$Gi=\frac{1}{2}(\gamma k_B T_c /H_c^2(0)\xi_{ab}^3(0))^2$
\cite{note_Gi} can be evaluated. While this dimensionless quantity
is of the order of $10^{-10}$ to $10^{-7}$ in low $T_{\text{c}}$
superconductors \cite{Forgan02,Eremenko02}, it can become larger
than $10^{-2}$ in H$T_{\text{c}}$SC cuprates \cite{Mikitik01}, due
to the higher $T_{\text{c}}$, the very short coherence length, and
the pronounced anisotropy. In the case of MgB$_2$, with parameters
we will obtain from $H_{\text{c2}}(T,\theta)$, we can estimate $Gi
\approx 10^{-5}$: The importance of fluctuations in MgB$_2$ is
halfway in between the one of fluctuations in H$T_{\text{c}}$SC
cuprates and the one of fluctuations in conventional low
$T_{\text{c}}$ compounds.

Fluctuations are difficult to tract theoretically in full
generality. In zero and low fields, fluctuations in high
$T_{\mathrm{c}}$ superconductors have been successfully described
with the ``3DXY model'' of critical fluctuations (see Schneider
and Singer \cite{Schneiderbook}, and references therein) In
sufficiently high magnetic fields, on the other hand, a different
approach, the so-called ``lowest Landau level'' (LLL)
approximation, was used successfully, both in high and low
$T_{\text{c}}$ compounds, to describe the effects of fluctuations
around $H_{\text{c2}}$
\cite{Tinkham_intro,Lee72,Ullah90,Welp91,Wilkin93}. The basic
physical idea \cite{Lee72} is that in a uniform field $H\|c$, the
fluctuating Cooper pairs are moving in quantized Landau orbits
characterized by $k_z$ and $n$. The mean field transition
temperature $T_{\text{c2}}(H)$ is the temperature, at which the
$n=0$ (i.e., the lowest) Landau level becomes stable, since Cooper
pairs do not exist in higher fields, in the mean field
approximation. Close to (and above) $T_{\text{c2}}(H)$, the lowest
Landau level is expected to dominate the fluctuation
contributions. For a field high enough, $H>H_{\text{LLL}}$, and
temperatures close to $T_{\text{c2}}(H)$, the approximation is
well fulfilled, but upon approaching, along $H_{\text{c2}}(T)$,
the zero field critical temperature $T_{\text{c}}$, the
contributions from Landau levels with $n>0$ can no longer be
neglected (see Lawrie \cite{Lawrie94} for a theoretical discussion
of the limits of the LLL approximation). At the same time, the
critical region is approached, and for low enough fields critical
fluctuations dominate.

%In the case of the H$T_{\text{c}}$SC cuprates, the values of
%$H_{\text{LLL}}$, and thus the field range where LLL and where
%critical fluctuations dominate, are a very controversial issue
%\cite{Lawrie94,Overend94,Pierson96,Jeandupeux96,Costa01}. In the
%case of MgB$_2$, we note that even with the theoretical criterion
%of Ref.\ \cite{Lawrie94}, which led to the high (in comparison
%with the findings of Refs.\ \cite{Pierson96,Jeandupeux96,Costa01})
%estimate $H_{\text{LLL}}\approx 200\,{\text{kOe}}$ in the case of
%YBa$_2$Cu$_3$O$_{7-\delta}$, we obtain an {\em upper limit}
%\cite{note_HLLLlimit} of $H_{\text{LLL}}\approx 5\,{\text{kOe}}$.
%Therefore, we expect a wider range of temperatures, where the LLL
%approximation should be valid near $H_{\text{c2}}(T)$ in MgB$_2$.
%For the analysis, it is also necessary to consider the
%dimensionality of the relevant fluctuations \cite{Wilkin93}. We
%are discussing the dimensionality of MgB$_2$ in some detail in
%Sec.\ \ref{lockin}, where we conclude that it is a three
%dimensional superconductor. Here, we only note that measurements
%of the fluctuation diamagnetism above $T_{\text{c}}$, in fields up
%to $5\,{\text{kOe}}$ \cite{Mosqueira02}, clearly indicate that
%fluctuations are of a three dimensional character in MgB$_2$.
In the case of MgB$_2$ \cite{Dulcic03}, we expect the LLL
approximation to be valid, except possibly very near
$T_{\mathrm{c}}$, and the fluctuations to be three dimensional in
nature \cite{AngstPhD}.
%For the $\theta_{\text{c2}}(H,T)$ analysis, we therefore assumed
%that fluctuations are three dimensional \cite{note_2Dflucttry} and
%that the LLL approximation is valid.
The fluctuation magnetization
$M$ of a 3D system in the vicinity of the transition temperature
$T_{\text{c2}}(H)$ in the LLL approximation is given by a
universal function F of $T-T_{\text{c2}}(H)$ \cite{Welp91}:
\begin{equation}
\frac{M}{H} = \frac{T^{2/3}}{H^{1/3}}\text{F}\left(
\frac{{\text{A}}(T-T_{\text{c2}})}{(TH)^{2/3}}\right),
\label{MHscal}
\end{equation}
where A is a material constant related to the Ginzburg number.
Eq.\ (\ref{MHscal}) applies for an isotropic superconductor as
well as for an anisotropic one. However, in the last case the
angular dependence of
$\text{F}({\text{A}}(T-T_{\text{c2}})/(TH)^{2/3})$ should be taken
into account. The equation can also be applied to the torque of an
anisotropic superconductor, by using a scaling transformation, as
done by Buzdin and Feinberg for arbitrary isotropic scaling laws
\cite{Buzdin92}. We find that the rescaled torque signal
\begin{equation}
\text{P}=-\tau\epsilon^{1/3}(\theta) \left/ \left(
\sin\theta\cos\theta H^{5/3}(1-1/\gamma_H^2)T^{2/3} \right)
\right.
 \label{scaling}
\end{equation}
with
$\epsilon(\theta)=(\cos^2\theta+\sin^2\theta/\gamma_H^2)^{1/2}$,
is, in the LLL approximation, a universal function of
$T-T_{\text{c2}}$ with a fixed value $\text{P}(0)$ at
$T=T_{\text{c2}}(H)$ \cite{note_scalingangle}.

\begin{figure}[tb]
\centering
\includegraphics[width=0.5\linewidth]{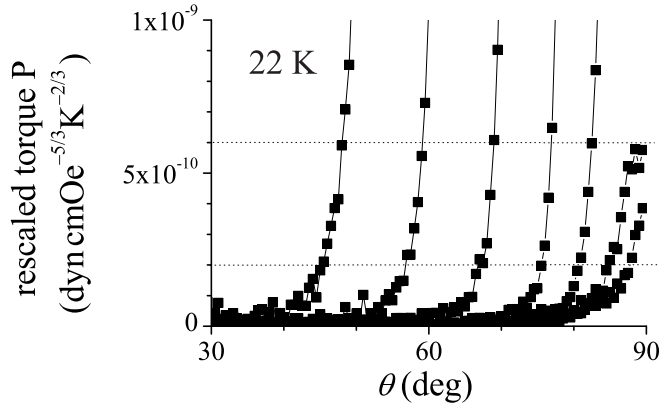}
\caption{ Selected rescaled torque P [see Eq.\ (\ref{scaling})] vs
angle $\theta$ curves measured on crystal B at $22\,{\mathrm{K}}$.
From left to right, curves shown were measured in $H=24$, $30$,
$40$, $55$, $70$, $80$, and $85\, \mathrm{kOe}$. Two criteria used
for the determination of $\theta_{\mathrm{c2}}$ are indicated by
horizontal dotted lines (see text). The $\theta_{\mathrm{c2}}$
values obtained employing the lower criterion are shown in Fig.\
\ref{Hc222K} below. After Ref.\ \cite{MgB2anisPRL02}.}
\label{Ptheta}
\end{figure}

\begin{figure}[tb]
\centering
\includegraphics[width=0.9\linewidth]{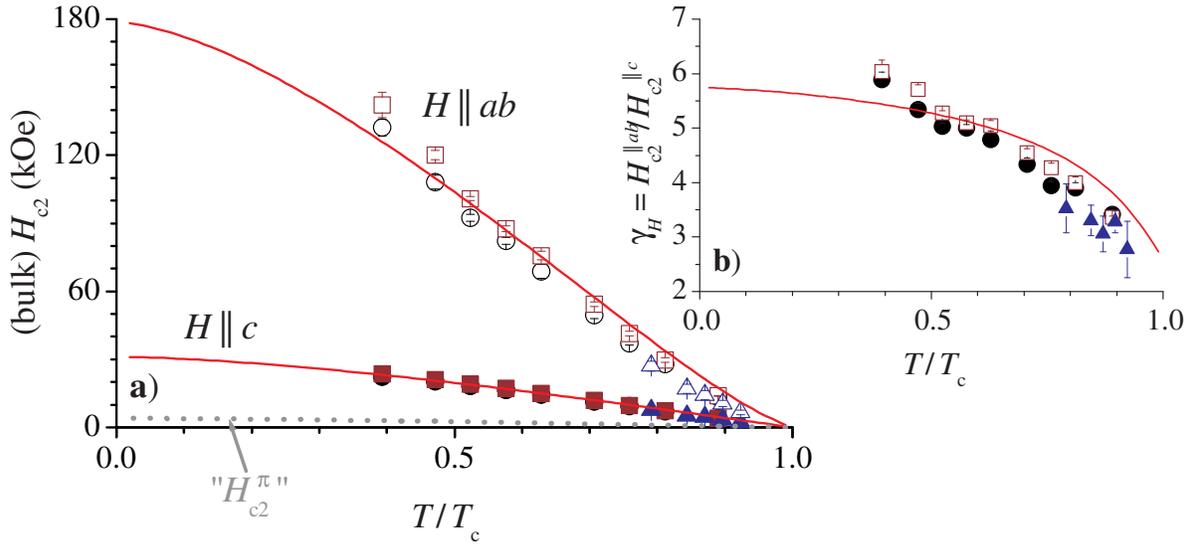}
\caption{ a) Upper critical field $H_{\mathrm{c2}}$ vs.\
temperature $T$. Open symbols correspond to $H\parallel ab$, full
symbols to $H\parallel c$, from fits of $H_{\mathrm{c2}}(\theta)$
data to Eq.\ (\ref{Hc2_theta}). Up triangles are from measurements
on sample A (with $\theta_{\mathrm{c2}}$ determined with a simple
``straight line crossing'' criterion) and squares (circles) are
from measurements on sample B, using two criteria of constant
rescaled torque P as defined in Eq.\ (\ref{scaling}). Full lines
are theoretical curves calculated by Miranovic {\em et al.}
\cite{Miranovic03} (see also text). The dashed line indicates
``$H_{\mathrm{c2}}^{\pi}$'', a crossover field above which
superconductivity in the $\pi$ bands is drastically suppressed. b)
Temperature dependence of the upper critical field anisotropy
$\gamma_H=H_{\mathrm{c2}}^{\|ab}/H_{\text{c2}}^{\|c}$, determined
from fits of $H_{\mathrm{c2}}(\theta)$ to Eq.\ (\ref{Hc2_theta}).
Symbol forms correspond to the same criteria/crystals as in panel
a). The full line is again from the theoretical calculation by
Miranovic {\em et al.} \cite{Miranovic03}. After Ref.\
\cite{Angst03MgB2PhyC}.} \label{gammaHvsT}
\end{figure}

Taking into account the $\text{F}(0)$ value for the theoretical
dependence of the universal function for a 3D system
\cite{Wilkin93}, we can estimate that for a volume of the sample
of $8\times 10^{-3}\, {\text{mm}}^3$ the rescaled torque P reaches
at $T=T_{\text{c2}}(H)$ a value of about $2\times 10^{-10}\,
{\text{dyn}}\, {\text{cmOe}}^{-5/3}{\text{K}}^{-2/3}$. Figure
\ref{Ptheta} shows a selection of torque vs angle curves rescaled
according to Eq.\ (\ref{scaling}), in different fields at
$22\,{\text{K}}$. The crossing of the ${\text{P}}(\theta)$
dependence for each field with the line of the constant value of
$2\times 10^{-10}\, {\text{dyn}}\,
{\text{cmOe}}^{-5/3}{\text{K}}^{-2/3}$ determines the
$H_{\text{c2}}(\theta)$ dependence as it is shown in Fig.\
\ref{Hc222K} below. Since there are some uncertainties in the
determination of the volume of a small crystal, we used for
comparison an additional criterion of $6\times 10^{-10}\,
{\text{dyn}}\, {\text{cmOe}}^{-5/3}{\text{K}}^{-2/3}$, i.e., three
times higher. The two criteria used are indicated by dashed lines
in Fig.\ \ref{Ptheta}. It is important to stress that the results
obtained depend not very sensitively on the criterion chosen and
it will be shown later (see Fig.\ \ref{gammaHvsT}) that with both
criteria defined here, we get very similar temperature dependences
of $H_{\text{c2}}$ and $\gamma$. Additional $\tau(H)$ measurements
at fixed angle give $H_{\text{c2}}(\theta)$ values corresponding
well to those from $\tau(\theta)$ measurements.

Since, for the purpose of the determination of
$H_{\text{c2}}(\theta)$, the transformation of the scaling
variable is not necessary, no full scaling analysis was performed.
From the resulting $H_{\text{c2}}(\theta)$ curve, the anisotropy
parameter $\gamma_H$ is then extracted by an analysis with AGLT,
i.e., Eq.\ (\ref{Hc2_theta}). We note that in the rescaling of the
torque according to Eq.\ (\ref{scaling}), the target parameter
$\gamma_H$ is used, which is obtained only later with Eq.\
(\ref{Hc2_theta}). Therefore, scaling analysis and determination
of $\gamma_H$ with Eq.\ (\ref{Hc2_theta}) had to be performed
iteratively in order to self consistently find $\gamma_H$.
However, the $\theta_{\text{c2}}(H)$ and $H_{\text{c2}}(\theta)$
points obtained with the scaling analysis depend not very strongly
on the value of $\gamma_H$ used in Eq.\ (\ref{scaling}), and the
iterative procedure converges rather fast.

\begin{figure}[tb]
\centering
\includegraphics[width=0.6\linewidth]{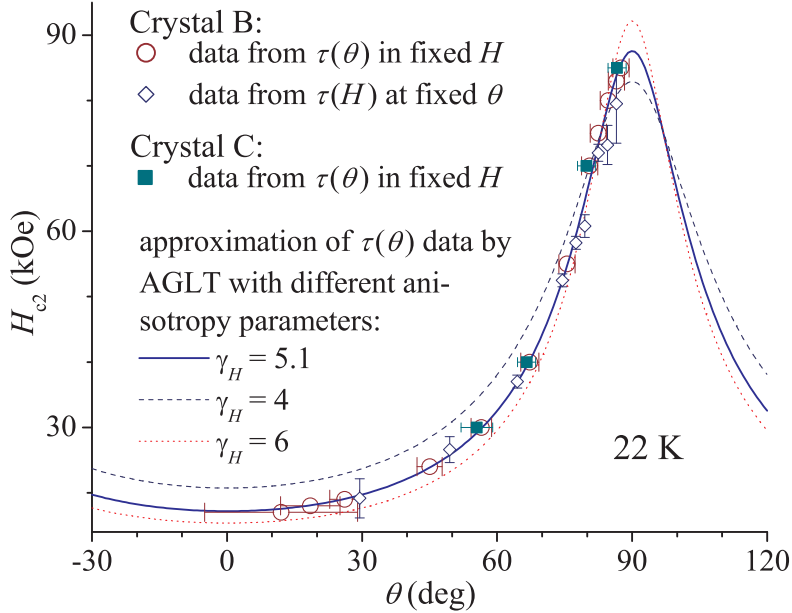}
\caption{ Upper critical field $H_{\text{c2}}$ vs angle $\theta$
at $22\,{\mathrm{K}}$. The data points were extracted with a
scaling analysis from $\tau(\theta)$ data measured in fixed field
($\circ$ from crystal B, $\blacksquare$ from crystal C, cf.\ Fig.\
\ref{Ptheta}) and from $\tau(H)$ data measured at fixed angle
($\diamond$), employing the lower of the two criteria described in
the text. Also shown are theoretical curves according to AGLT,
with anisotropy parameters $\gamma_H=5.1$ (full curve), $4$
(dashed curve), and $6$ (dotted curve). After Refs.\
\cite{MgB2anisPRL02,Angst03MgB2PhyC}.} \label{Hc222K}
\end{figure}

Figure \ref{Hc222K}, where all $\theta_{\text{c2}}(H)$ and
$H_{\text{c2}}(\theta)$ points, obtained from the scaling analysis
of the data measured at $22\,{\text{K}}$ on crystal B, are
plotted, shows that the data obtained employing the LLL scaling
analysis are very well described by the AGLT formula [Eq.\
(\ref{Hc2_theta})], with an anisotropy parameter $\gamma_H=5.1$.
Alternative curves from fits with Eq.\ (\ref{Hc2_theta}), but
using $\gamma_H=4$ or $6$, clearly fail to describe the data. The
figure also shows data points obtained from crystal C
\cite{note_anisC}. The irreversible properties of crystals B and C
are rather different, i.e., they have a different defect
structure. The good agreement both in value and angular dependence
of $H_{\mathrm{c2}}$ of crystals B and C that is observable in
Fig.\ \ref{Hc222K} indicates that such differences in the defect
structure do not influence the upper critical field much, at least
in the region between $22$ and $34\,{\mathrm{K}}$
\cite{note_anisC}, and therefore cannot influence our conclusion
of a $T$ dependent $H_{\mathrm{c2}}$ anisotropy. We found the
angular dependence of $H_{\text{c2}}$ to be well described by Eq.\
(\ref{Hc2_theta}) at all temperatures not too close to
$T_{\mathrm{c}}$, but with a temperature dependent anisotropy
parameter $\gamma_H(T)$.

The first theoretical $H_{\mathrm{c2}}(T,\theta)$ calculation
based on a two band model (using the clean limit, i.e., ignoring
both inter- and intraband scattering) \cite{Miranovic03} confirmed
the temperature dependent $\gamma_H(T)$, but indicated systematic
deviations of the angular $H_{\mathrm{c2}}$ dependence more
pronounced at low temperatures \cite{Miranovic_pc}, predicting
that the AGLT angular dependence should be followed well near
$T_{\mathrm{c}}$. This is at variance with the experimental
results, as we will discuss now.

\begin{figure}[tb]
\centering
\includegraphics[width=0.85\linewidth]{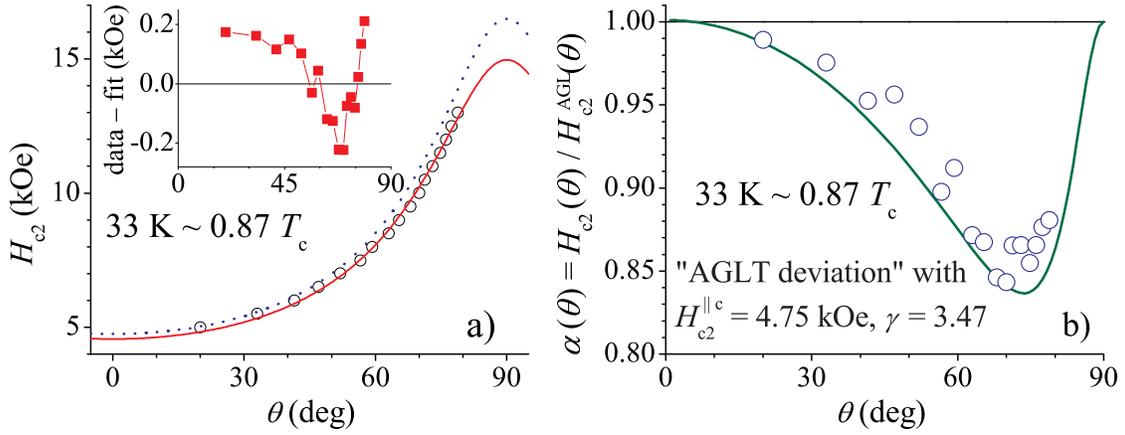}
\caption{ a) Upper critical field $H_{c2}$ vs angle $\theta$ of a
MgB$_2$ single crystal (A), at $0.87\,T_c$ (symbols). Free fit of
AGLT angular dependence $H_{\text{c2}}^{\mathrm{AGL}}(\theta)$
[Eq.\ (\ref{Hc2_theta})] (full line;
$H_{c2}^{\|c}=4.56\,{\mathrm{kOe}}$, $\gamma _H = 3.28$) shows
clear systematic deviations, highlighted in the inset. AGLT
dependence with the same parameters as used in panel b) is also
shown (dotted line, $H_{c2}^{\|c}=4.75\,{\mathrm{kOe}}$, $\gamma
_H = 3.47$). b) "AGLT deviation"
$\alpha(\theta)=H_{\text{c2}}(\theta)/H_{\text{c2}}^{\mathrm{AGL}}(\theta)$
(see text) of the data of panel a) vs angle $\theta$. Full line:
calculation of Golubov and Koshelev \cite{Golubov03}, for the same
reduced temperature. After Ref.\ \cite{Angst03Rio}.}
\label{AGLTdev}
\end{figure}

The upper critical field $H_{c2}(\theta)$, determined from
$\tau(\theta)$ curves measured in various fields at
$33\,{\mathrm{K}} \simeq 0.87\, T_c$, is shown in Fig.\
\ref{AGLTdev}a). By definition, $\tau$ is 0 for $H\|c$ and
$H\|ab$, and small for field directions close. This is why there
are no data close to $0^{\circ}$ and $90^{\circ}$. The best fit of
Eq.\ (\ref{Hc2_theta}) to the data is indicated by the full line.
Small, but systematic deviations can be seen, especially when
plotting the difference between experimental data and best fit vs
$\theta$ (inset): at $0.87\, T_c$, $H_{c2}(\theta)$ is {\em not}
(accurately) described by Eq.\ (\ref{Hc2_theta}). Deviations from
Eq.\ (\ref{Hc2_theta}) were not observed at lower $T$ (cf.\ Fig.\
\ref{Hc222K}). Of course, in principle our experimental limitation
of fields up to $90\,{\text{kOe}}$ may prevent the observation of
deviations from Eq.\ (\ref{Hc2_theta}) at low temperatures
($\lesssim 20\,{\mathrm{K}}$). On the other hand, an independent
investigation \cite{Lyard02} of $H_{\text{c2}}(\theta)$,
determined from electrical transport measurements
\cite{note_resistLyard} in fields up to $280\,{\text{kOe}}$, found
no deviations of the angular $H_{\text{c2}}$ dependence from the
prediction of Eq.\ (\ref{Hc2_theta}), even at $5\,{\text{K}}$.

Deviations most pronounced in the region of $0.9$-$0.95\, T_c$
were also found in a recent calculation \cite{Golubov03} assuming
zero interband, but high intraband impurity scattering ({\em dirty
limit}, see also Gurevich \cite{Gurevich03}). In order to compare
the deviations observed in torque experiments to the predictions
of Golubov and Koshelev \cite{Golubov03}, we calculated ``AGLT
deviations'' $\alpha (\theta) \equiv H_{c2} (\theta) /
H_{c2}^{\mathrm{AGL}} (\theta)$. For $H_{c2}^{\|c}=
4.75\,{\mathrm{kOe}}$ and $\gamma_H = 3.47$, $\alpha (\theta)$ has
form and magnitude [Fig.\ \ref{AGLTdev}b)] very similar to
deviation functions for {\em calculated} \cite{Golubov03}
$H_{c2}(\theta)$ (full line) at the same temperature \cite{note2}.
Although the theoretically predicted \cite{Golubov03} $\gamma_H
\simeq 4.86$ is higher than our data indicate, the similarity of
the AGLT deviation suggests that (intraband) impurity scattering
cannot be neglected in theoretical descriptions of $H_{c2}$ in
MgB$_2$, even in single crystals of highest quality.

The upper critical fields parallel and perpendicular to the layers
obtained with the scaling analysis and Eq.\ (\ref{Hc2_theta}) are
shown in Fig.\ \ref{gammaHvsT}a). Results obtained from two
crystals in two magnetometers and using three different
$H_{\text{c2}}$ criteria are depicted as different symbols. It can
be seen that the values of $H_{\text{c2}}^{\|c}(T)$ and
$H_{\text{c2}}^{\|ab}(T)$ do not depend very sensitively on the
exact $H_{\text{c2}}$ criterion used. The $T$ dependence of
$H_{\text{c2}}^{\|c}$ is in agreement with (isotropic)
calculations by Helfand and Werthamer \cite{Helfand66}, with
$H_{\text{c2}}^{\|c}(0)\simeq 31\, {\text{kOe}}$. On the other
hand, $H_{\mathrm{c2}}^{\|ab}(T)$ exhibits a slight positive
curvature near $T_{\mathrm{c}}$ (due to the lack of low
temperature data, for $H_{\text{c2}}^{\|ab}(0)$ only an estimation
$180\, {\text{kOe}}\lesssim H_{\text{c2}}^{\|ab}(0)\lesssim 230\,
{\text{kOe}}$ can be given). The positive curvature of
$H_{\mathrm{c2}}^{\|ab}(T)$ may also account for the positive
curvature of $H_{\mathrm{c2}}$ observed in earlier measurements of
polycrystalline material \cite{Buzea01}. Correspondingly, the
anisotropy $\gamma_H=H_{\text{c2}}^{\|ab}/H_{\text{c2}}^{\|c}$
{\em{systematically}} {\em{decreases}} with increasing temperature
[Fig.\ \ref{gammaHvsT}b)], as we already expected from Fig.\
\ref{rawcurves}. A change of the criterion used for the
determination of $\theta_{\text{c2}}$ leads to small shifts of the
magnitude of $\gamma_H$, but the temperature dependence is always
the same. The highest upper critical field anisotropy
$\gamma_H\simeq 6$ was obtained at $15\, {\text{K}}$, the lowest
anisotropy $\gamma_H\simeq 2.8$ at $35\, {\text{K}}$. From Fig.\
\ref{gammaHvsT} we estimate $\gamma_H(T_{\text{c}}) = 2.3-2.7$,
while at zero temperature, $\gamma_H$ may become as large as $8$.

Since the time our torque measurements were performed, many other
authors also observed a temperature dependence of $\gamma_H$ in
MgB$_2$ as well
\cite{Eltsev02,Ferdeghini02,Sologubenko02,Budko02,Zehetmayer02,Lyard02,Welp03,Machida03,Cooper_pc}.
Before making any quantitative comparisons, we note that
electrical transport measurements \cite{Eltsev02,Ferdeghini02}
yield too high values of $H_{\text{c2}}^{\|c}$, due to surface
superconductivity (``$H_{\text{c3}}$'') \cite{Welp03} or more
complicated effects \cite{Sologubenko02}. All bulk measurements
(torque \cite{MgB2anisPRL02,Cooper_pc}, magnetization
\cite{Budko02,Zehetmayer02,Welp03,Machida03}, thermal conductivity
\cite{Sologubenko02}, and specific heat \cite{Lyard02,Welp03})
agree well on the $H_{\text{c2}}^{\|c}(T)$ dependence and value.
Concerning $H_{\text{c2}}^{\|ab}(T)$, and consequently
$\gamma_H(T)$, however, reported values differ from each other. It
is difficult to say exactly what part of the discrepancies is to
be attributed to the difference in experimental methods and
procedures, and what part originates from the different samples
used. The deviations of the angular dependence of
$H_{\mathrm{c2}}$ from Eq.\ (\ref{Hc2_theta}) near
$T_{\mathrm{c}}$ lead to a slight underestimation of
$H_{\mathrm{c2}}^{\|ab}/H_{\text{c2}}^{\|c}$ from our torque
measurements in this region, most pronounced for $T\simeq
0.9-0.95\, T_{\mathrm{c}}$. This may explain at least partly the
distribution over a wider temperature range of the change in
anisotropy compared to experiments measuring directly
$H_{\mathrm{c2}}^{\|ab}(T)$ and $H_{\mathrm{c2}}^{\|c}(T)$. This
problem should not affect the data at lower temperature, where the
values of $H_{\text{c2}}^{\|ab}(T)$ and $\gamma_H(T)$ obtained in
this work is closer to the upper end of the spread of experimental
reports. We note that the limited field range ($\lesssim
90\,{\text{kOe}}$) in the experiment inevitably reduced the
accuracy of our analysis in the case of the lowest temperature
data shown in Fig.\ \ref{gammaHvsT}, but since no systematic
deviations from Eq.\ (\ref{Hc2_theta}) are indicated in our and
other measurements, we think the points shown are still reliable.
It is possible that a small number of stacking faults may affect
the $H_{\text{c2}}^{\|ab}$ value in this region, as discussed by
Angst \cite{AngstPhD}. However, a comparison of
$H_{\text{c2}}(\theta ,T)$ of crystals from the same source, but
grown with slightly varied procedures and pronouncedly different
pinning properties yielded no significant differences (cf.\ Fig.\
\ref{Hc222K} and Angst {\em et al.} \cite{Angst03MgB2PhyC}). On
the other hand, the low temperature upper critical field of
crystals from different sources can be more different, which may
be due to significantly different intraband impurity scattering
rates in the $\sigma$ and the $\pi$ band \cite{Carrington03}.

The anisotropic upper critical field of MgB$_2$ has been
calculated by several authors using different approaches
\cite{Posazhennikova02,Miranovic03,Gurevich03,Dahm02,Golubov03}.
Apart from the report by Posazhennikova {\em et al.}
\cite{Posazhennikova02}, which was based on an anisotropic $s$
wave model, all calculations were based on the two band model (see
Sec.\ \ref{twoband}). The calculations agree qualitatively with
each other as well as with experiments. For comparison, the curves
calculated by Miranovic {\em at al.} \cite{Miranovic03} are also
shown in Fig.\ \ref{gammaHvsT}.
%The values of
%$H_{\text{c2}}^{\|ab}(T)$ and $\gamma_H(T)$ obtained in this work
%is closer to the upper end of the spread. If the differences
%between reports is mainly due to sample differences, we speculate
%that in our sample B \cite{note_ABC,note_AC}, a small number of
%stacking faults may reduce the effective coherence length
%perpendicular to the planes \cite{AngstPhD} and hence increase
%$H_{\text{c2}}^{\|ab}(T)$, particularly at low temperatures.
%
Concerning the details, however, there are variations between the
theoretical reports, as in the case of the experimental ones. For
example, using similar input values from band structure and
electron phonon coupling calculations (cf.\ Sec.\ \ref{twoband}),
values of $\gamma_H (0)$ of $5.7$ \cite{Miranovic03}, $4.6$
\cite{Dahm02} and about $6.1$ \cite{Golubov03} were obtained.

A major difference between the calculations of Refs.\
\cite{Miranovic03,Dahm02} and the ones of Refs.\
\cite{Gurevich03,Golubov03} is that the former neglect all
impurity scattering, while the latter use an approach
\cite{Usadel70} based on the $\sigma$ and the $\pi$ band being in
the ({\em intraband}) dirty limit. It was demonstrated that in
this case the upper critical fields, their anisotropy, and the
temperature dependence of these quantities depend sensitively on
the intraband scattering rates in the $\pi$ and in the $\sigma$
band \cite{Gurevich03}. Indications that this may indeed the case
in MgB$_2$ are given for example by a much lower anisotropy with a
decreased temperature dependence in carbon substituted MgB$_2$
\cite{Ribeiro03} and also by the wide spread of values of
$H_{\mathrm{c2}}$ in various samples in early reports
\cite{Buzea01}. Furthermore, the differences seen even in the case
of high quality single crystals \cite{Carrington03} and the
agreement of experimental results with the calculations by Golubov
and Koshelev \cite{Golubov03} concerning the deviations of the
angular dependence of $H_{\mathrm{c2}}$ from Eq.\
(\ref{Hc2_theta}) (see above) strongly suggest that intraband
scattering is important in all present samples of MgB$_2$. We note
that there remains the problem that experiments, e.g.\ de Haas-van
Alphen \cite{Yelland02,Carrington03}, magneto-optics
\cite{Perucchi02}, and scanning tunnelling spectroscopy
\cite{Eskildsen02} indicate the $\pi$ bands to be in the
(moderately) dirty limit, but the $\sigma$ bands to be rather in
the (moderately) clean limit whereas the comparison to experiments
of the calculations carried out in the dirty limit indicates
higher scattering rates in the $\sigma$ bands \cite{Koshelev03}.

Physically, it seems clear from a comparison of Fig.\
\ref{gammaHvsT} and the electronic anisotropy in the $\sigma$
bands ($\sim 6.8^2$, see Sec.\ \ref{twoband}) that the low
temperature upper critical field and its anisotropy is dominated
by these bands with the stronger superconductivity (larger gap).
In the hypothetical situation of no coupling between $\sigma$ and
$\pi$ bands, there would be two different upper critical fields
for these bands with $H_{\mathrm{c2}}^{\pi}$ nearly isotropic and
much smaller than $H_{\mathrm{c2}}^{\sigma}$, due to the smaller
superconducting gap in the $\pi$ bands [cf.\ Eqs.\ (\ref{xigap})
and (\ref{xiHc2})]. Due to the coupling by interband pairing and
quasiparticle exchange, $H_{\mathrm{c2}}^{\pi}$ becomes a mere
crossover field \cite{Bouquet02} with superconductivity in the
$\pi$ bands induced by the $\sigma$ bands superconductivity
\cite{Eskildsen02} even above this field. This crossover field is
indicated in Fig.\ \ref{gammaHvsT} by a dashed line. This issue is
discussed in more detail in Sec.\ \ref{mixedstate}. The $\pi$
bands retain some influence on the bulk upper critical field due
to coupling, but at low temperature this influence is not very
large for the small interband pairing strength characteristical
for MgB$_2$ \cite{Dahm02}.

In the limit $T\rightarrow T_{\mathrm{c}}$, on the other hand,
thermal excitations lead to a thorough mixing of states over the
whole Fermi surface, and the microscopic derivation of the
anisotropic Ginzburg-Landau equations \cite{Gorkov64} is expected
to hold, even if this is not the case at lower temperatures (cf.\
Ref.\ \cite{AngstPhD}). This leads to an anisotropy parameter
being a simple average over the Fermi surface of the electronic
anisotropy as a function of the wave vector, either weighted by
the gap value (clean case, $\sim 2.6$ in MgB$_2$) or evenly
weighted (dirty case, $\sim 1.2$ in MgB$_2$)
\cite{Gorkov64,Kogan02,Miranovic03}. The pronounced temperature
dependence of the upper critical field anisotropy of MgB$_2$ is,
then, a crossover between the above two cases.

Although anisotropic Ginzburg-Landau theory should hold for
$T\rightarrow T_{\mathrm{c}}$, Fig.\ \ref{gammaHvsT} clearly
indicates that the variation of $\gamma_H$ with temperature is the
strongest close to $T_{\text{c}}$. Furthermore, also the
deviations of the angular dependence of $H_{\mathrm{c2}}$ from
AGLT predictions is observed to be particularly strong near
$T_{\mathrm{c}}$ [see Fig.\ \ref{AGLTdev}]. Therefore, in MgB$_2$,
AGLT seems to have an exceptionally limited range of applicability
indeed. According to the calculations by Golubov and Koshelev
\cite{Golubov03}, the AGLT description is indeed {\em limited to
temperatures less than $2\%$ away from $T_{\mathrm{c}}$}. Since
such a small temperature region is difficult to accurately probe
experimentally, one may indeed say that there is a complete {\em
breakdown of anisotropic Ginzburg-Landau theory} in MgB$_2$.

In this small AGLT region very close to $T_{\mathrm{c}}$, Eq.\
(\ref{gammas}) should hold, i.e.\ with the definition
$\gamma_{\lambda}\equiv \lambda_c / \lambda_{ab}$, we should have
$\gamma_H(T_{\text{c}})=\gamma_{\lambda}(T_c)\simeq 2.6$. At lower
temperatures, where AGLT no longer holds, however, there is no
reason why $\gamma_{\lambda} (T)$ should be equal to $\gamma_H
(T)$. Indeed, calculations of $\gamma_{\lambda} (T)$ ({\em in the
limit of very low fields}) \cite{Kogan02,Golubov02b} found the
penetration depth anisotropy to increase with increasing
temperature, from close to $1$ at $0\,{\text{K}}$ to the ``AGLT
anisotropy'' at $T_{\mathrm{c}}$, exactly opposite to the behavior
of $\gamma_H (T)$. Furthermore, Golubov \cite{Golubov02b}
predicted a strong deviation of the angular scaling of the
penetration depth from the AGLT dependence [Eq.\
(\ref{Hc2_theta})], essentially due to the very small $\sigma$
bands $c$ axis plasma frequency \cite{Brinkman02}. This has not
yet been experimentally confirmed, but as we have seen above, such
deviations have been found in the case of the upper critical
field. The so far only report on direct measurements of the low
field penetration depth on single crystals \cite{Manzano02} uses a
method that is not very sensitive to the absolute anisotropy value
at zero temperature, but concludes that there is very little
temperature dependence of $\gamma_ \lambda$. This clearly implies
$\gamma_ \lambda \neq \gamma_H$ for most temperatures. Such a
difference of anisotropies should influence the mixed state
properties, particularly the torque far below $H_{\text{c2}}$.

Indeed, the field delimiting the mixed state from below, the lower
critical field $H_{\mathrm{c1}}$, is mainly controlled by the
penetration depth. From the energy needed to create an isolated
vortex, it follows (within anisotropic Ginzburg-Landau theory)
$H_{\mathrm{c1}}^{\|c}= \Phi _{\circ} / 4 \pi \lambda _{ab}^2
\times \ln (\lambda _{ab}/\xi_{ab})$. Early measurements of the
lower critical field of MgB$_2$, performed mostly on
polycrystalline material, yielded values between $250$ and
$480\,{\mathrm{Oe}}$. Meanwhile a few measurements on single
crystals were reported. A direct measurement on single crystal by
SQUID magnetometry gave values of $H_{\mathrm{c1}}^{\|c}(0)\simeq
380\,{\mathrm{Oe}}$ and $H_{\mathrm{c1}}^{\|ab}(0)\simeq
270\,{\mathrm{Oe}}$, with an anisotropy of $1.4$ \cite{Xu01}. A
later direct measurement with a vibrating sample magnetometer
yielded the surprisingly high values of
$H_{\mathrm{c1}}^{\|c}(0)\simeq 2800\,{\mathrm{Oe}}$,
$H_{\mathrm{c1}}^{\|ab}(0)\simeq 1350\,{\mathrm{Oe}}$, with an
anisotropy of $2.1$ \cite{Perkins02}. Finally, direct measurements
of $H_{\mathrm{c1}}^{\|c}(5\,{\mathrm{K}})\simeq
700\,{\mathrm{Oe}}$ $H_{\mathrm{c1}}^{\|ab}(0)\simeq
220\,{\mathrm{Oe}}$ were reported \cite{Zehetmayer02}. It should
be noted that direct measurements of the lower critical field of
single crystals are very difficult, because (i) only the
penetration field $H_{\mathrm{p}}$, where the first vortex enters
the sample, is obtained, which can be higher than
$H_{\mathrm{c1}}$ for example due to geometrical barriers
\cite{Zeldov94}, (ii) demagnetizing effects lead to an
underestimation of $H_{\mathrm{p}}$ if not taken into account
properly, particularly in thin crystals for $H\|c$. From our
torque measurements, we obtain at low temperatures for different
samples values in the region of $H_{\mathrm{p}}\approx
350-500\,{\mathrm{Oe}}$, with little dependence on the field
direction. This is compatible with the early measurements and
would be compatible with a low penetration depth anisotropy
\cite{Kogan02,Golubov02b}. However, in light of the difficulties
associated with the direct determination of $H_{\mathrm{c1}}$ and
the wide spread of values reported for single crystals, we have to
conclude that the situation regarding the lower critical field of
MgB$_2$ is much less clear than the situation regarding the upper
critical field. For example there is no report about the angular
dependence of $H_{\mathrm{c1}}$, where deviations from the AGLT
description may occur like in the case of $H_{\mathrm{c2}}$
discussed above. Often the values of $H_{\mathrm{c1}}$, as well as
those of $H_{\mathrm{c2}}$ are deduced indirectly from the
magnetization in the mixed state, for example on the basis of the
London model (see, e.g., Zehetmayer {\em et al.}
\cite{Zehetmayer02}). The mixed state $H_{\mathrm{c1}} < H <
H_{\mathrm{c2}}$ is discussed in the next section, Sec.\
\ref{mixedstate}.

\section{Anisotropies for $H < H_{\mathrm{c2}}$: there are still open questions, but ``the smaller the field the more isotropic''}

\label{mixedstate}

To obtain information about the anisotropy (and $\lambda$ and
$\xi$) in the mixed state $H_{\text{c1}} \lesssim H \ll
H_{\text{c2}}$ (existing in superconductors with high enough
$\kappa = \lambda / \xi$), an analysis based on the ``London
model'' \cite{Tinkham_intro} was often used. This method was for
example used with great success in the case of the high
$T_{\mathrm{c}}$ cuprate superconductors (see, e.g., Refs.\
\cite{Farrell88,Farrell89,Zech96,Willemin98,Hofer98}), where the
direct measurement of the $H_{\mathrm{c2}}$ anisotropy is very
difficult due to the blurring out of the transition by
fluctuations (cf.\ Sec.\ \ref{Hc2}).

The {\em basic} assumption of London model is not high $\kappa$
and $H\ll H_{c2}$, but a constant magnitude of the order parameter
throughout most of the space. Only in this case the variation of
supercurrents and local magnetization is ``simple'' in most of the
region and the small rest of the region (essentially the cores)
can be treated in semi-phenomenological cut-off. This gets
translated to ``no core overlap'', restricting the London regime
to fields sufficiently below $H_{c2}$. In the London limit,
$H_{\text{c1}}\ll H \ll H_{\text{c2}}$ and $\kappa\gg 1$,
calculations by Kogan and coworkers
\cite{Kogan81,Kogan88,Kogan88b} of the magnetization components
longitudinal and perpendicular to the field lead to a torque given
by \cite{Zech96}
\begin{equation}
\tau = \frac{V \Phi_{\circ}H}{64 \pi^2 \lambda_{ab}^2} \left ( 1 -
\frac{1}{\gamma_{\mathrm{eff}}^2} \right ) \frac{\sin 2
\theta}{\epsilon(\theta)} \ln \left ( {\frac{\eta
H_{\text{c2}}^{\|c}}{\epsilon (\theta) H}} \right ),
\label{tau_rev}
\end{equation}
where $\epsilon(\theta)=( \cos^2 \theta + \sin^2 \theta /
\gamma_{\mathrm{eff}}^2 )^{1/2}$, $V$ is the sample volume, and
$\gamma_{\mathrm{eff}}$ is an effective anisotropy, equal to the
upper critical field anisotropy if Eq.\ (\ref{gammas}) is valid.
We note that Eq.\ (\ref{tau_rev}) is only approximately valid,
since the London approach ignores the contribution of the vortex
cores. From studying a more elaborate variational model it was,
however, found that Eq.\ (\ref{tau_rev}) still is a good
approximation in a range of intermediate fields, although with
renormalized parameter $\eta$ and renormalized penetration depth
\cite{Hao_Clem91,Hao91b}.

On the other hand, Eq.\ (\ref{tau_rev}) describes the reversible
equilibrium torque, whereas in low fields and at low temperatures,
the torque usually contains an irreversible part, that depends, on
the direction of the field change, or in constant field, on the
direction of the change of the angle between field and $c$ axis of
the sample. The irreversible part is caused by pinning due to
unavoidable point-like defects, due to extended defects such as
stacking faults, and in highly anisotropic materials even due to
the layered structure itself (see Sec.\ \ref{PE} for a discussion
of irreversible properties).

\begin{figure}[bt]
\centering
\includegraphics[width=0.9\linewidth]{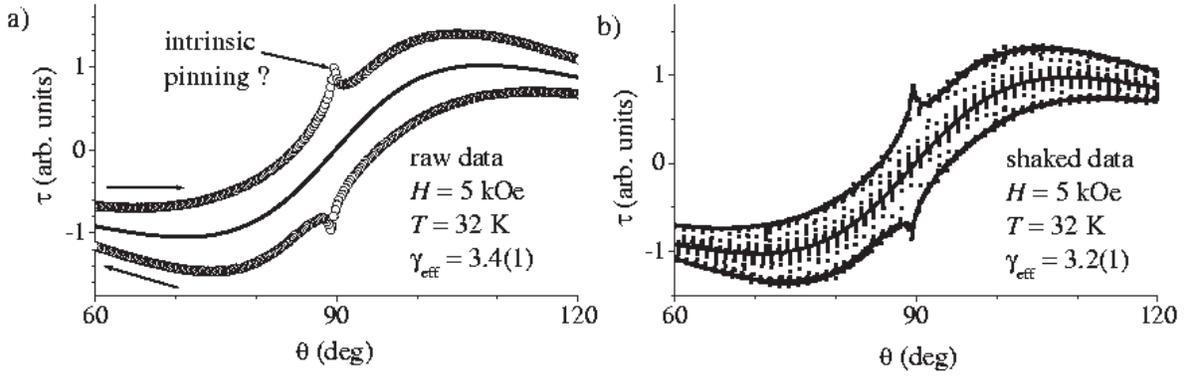}
\caption{ Torque $\tau$ of MgB$_2$ single crystal A vs angle
$\theta$ at $32 \, {\mathrm{K}}$ in $5\,{\mathrm{kOe}}$. a) Raw
data before shaking (open circles). The full line is a fit of Eq.\
(\ref{tau_rev}) to the average of data collected with angle
$\theta$ increasing and decreasing. For
$\theta=90\,{\mathrm{deg}}$, a peak in the irreversible torque can
be seen (see text). b) Raw data collected with a ``shaking''
procedure (full squares). The full line is a fit of Eq.\
(\ref{tau_rev}) to the ``shaked data''. After Refs.\
\cite{MgB2anisPRL02,CommentTakahashi}.} \label{tauthetashake}
\end{figure}

The combination of these effects significantly restricts the
region where Eq.\ (\ref{tau_rev}) describes measured torque data.
In fact, it is quite possible to have a material where there is no
region where Eq.\ (\ref{tau_rev}) is strictly applicable. In the
past, the equilibrium torque was often approximated by just
arithmetically averaging the measured torque for $H$/$\theta$
increasing and decreasing. However, such a procedure was shown
\cite{Willemin98} to lead to systematic errors in the deduction of
the anisotropy. A better way to obtain the true equilibrium torque
is to speed up the relaxation of the vortex lattice into it's
reversible state by an additional small oscillating magnetic field
perpendicular to the main field (``vortex shaking'')
\cite{Willemin98}. Note that this ``shaking'' procedure is
different from performing minor hysteresis loops (cf.\ Sec.\
\ref{PE}). A detailed theoretical explanation of this speed up of
the relaxation to the equilibrium by the shaking procedure was
recently given by Brandt and Mikitik \cite{note_Brandtshake}.

Figure \ref{tauthetashake} shows an example of one of the
$\tau(\theta)$ curves measured on crystal A for an analysis with
Eq.\ (\ref{tau_rev}). The comparison of panels a) and b)
demonstrates the importance of the vortex shaking technique for an
accurate analysis. Whereas the curves obtained without shaking
[panel a)] show a substantial degree of irreversibility, this
irreversibility is removed after shaking [panel b)] and the data
are well described by Eq.\ (\ref{tau_rev}), with an anisotropy
parameter $\gamma_{\mathrm{eff}}=3.2$. The ``unshaked'' data of
panel a), on the other hand, can be described by Eq.\
(\ref{tau_rev}) as well, but with a slightly higher least-squares
deviation, and with a higher anisotropy parameter,
$\gamma_{\mathrm{eff}}=3.4$. Although the difference in
$\gamma_{\text{eff}}$ is not very large and by the eye the fitting
curves in the two panels are difficult to distinguish, the data
seem to indicate that in MgB$_2$, the presence of irreversibility
leads to an overestimation of the anisotropy
\cite{CommentTakahashi}, similar to earlier findings in the case
of YBa$_2$Cu$_3$O$_{7-\delta}$ \cite{Willemin98}. In Fig.\
\ref{tauthetashake}a), a pronounced peak is visible in the
irreversible torque for $\theta=90\,{\text{deg}}$, reminiscent of
the well known ``intrinsic pinning'' of highly anisotropic cuprate
superconductors. This feature is discussed by Angst and coauthors
\cite{Karpinski03SST,AngstPhD}.

\begin{figure}[tb]
\centering
\includegraphics[width=0.5\linewidth]{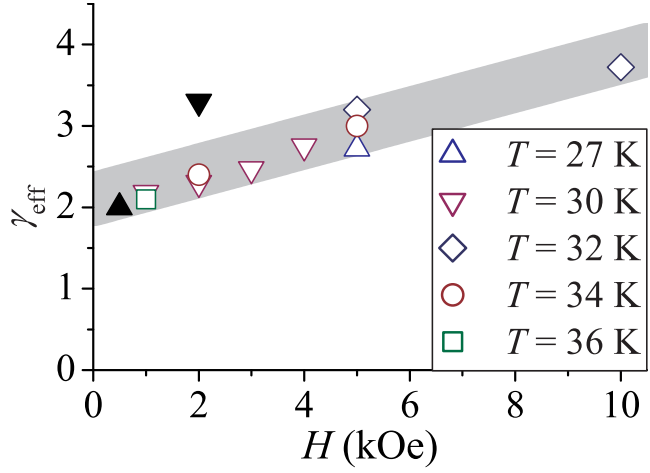}
\caption{ Field dependence of the effective anisotropy
$\gamma_{\mathrm{eff}}$ (open symbols,see text), obtained from an
analysis of reversible torque $\tau(\theta)$ data measured on
crystal A with Eq.\ (\ref{tau_rev}). After Ref.\
\cite{MgB2anisPRL02}. Also shown are two anisotropies determined
(see Fig.\ \ref{taurevlowT}) on crystal D at $8\,{\mathrm{K}}$
($\blacktriangledown$) and $37\,{\mathrm{K}}$ ($\blacktriangle$).}
\label{gammaeffH}
\end{figure}

A number of additional torque vs angle curves were measured on
crystal A not far away from $T_{\mathrm{c}}$ in relatively low
fields. Checking first the applicability of Eq.\ (\ref{tau_rev}),
with respect to the limitations sketched above, we note first that
all data were measured well above $H_{\text{c1}}$ and that the
curves were well reversible after shaking \cite{note_smallirr}. We
believe that due to the low Ginzburg number of MgB$_2$ (as
compared to the cuprates), critical fluctuations should not
appreciably influence the torque in the region of temperatures
measured. The remaining limitation is the influence of
$H_{\text{c2}}$ fluctuations and vortex core contributions in high
fields. The upper field limitation of the applicability of Eq.\
(\ref{tau_rev}) is not any well defined line in the $H$-$T$
diagram, but a comparison with the results of Hao and Clem
\cite{Hao91b} indicates that for the data shown below (Fig.\
\ref{gammaeffH}), Eq.\ (\ref{tau_rev}) should be a good
approximation, with the possible exceptions of the points
$H=10\,{\text{kOe}}$/$T=32\,{\text{K}}$ and
$H=5\,{\text{kOe}}$/$T=34\,{\text{K}}$.

The result of the analysis of reversible torque $\tau(\theta)$
data, measured at on crystal A at temperatures between $27$ and
$36\,{\text{K}}$ in fields from $1$ to $10\,{\text{kOe}}$, with
Eq.\ (\ref{tau_rev}) \cite{MgB2anisPRL02} is presented in Fig.\
\ref{gammaeffH} (open symbols). There is a number of points worth
noting:
\begin{itemize}
 \item The effective anisotropy $\gamma_{\text{eff}}$ is field
 dependent, increasing nearly linearly from $\gamma_{\text{eff}}\simeq
 2$ in zero field to $3.7$ in $10\,{\text{kOe}}$, again in disagreement
 with the anisotropic Ginzburg Landau theory.
 \item No clear temperature dependence of $\gamma_{\text{eff}}$ is visible in the range of
 temperatures of $27-36\,{\mathrm{K}}$ and fields below $10\,{\mathrm{kOe}}$.
 \item The effective anisotropy $\gamma_{\text{eff}}$ obtained
 from the analysis with Eq.\ (\ref{tau_rev}) is different from the
 upper critical field anisotropy $\gamma_H$ (see Sec.\ \ref{Hc2}).
\end{itemize}

Before proceeding with the discussion of these results, we compare
with other torque measurements analyzed with Eq.\ (\ref{tau_rev})
that were reported in the literature. One report of an analysis of
torque data measured at $10\,{\text{K}}$ on a small MgB$_2$
crystal with Eq.\ (\ref{tau_rev}) concluded that in that case,
$\gamma_{\text{eff}}$ was field independent \cite{Takahashi02}.
However, much of the data used in that analysis clearly is
measured in fields above the applicability region of Eq.\
(\ref{tau_rev}), partly even above $H_{\mathrm{c2}}^{\|c}$, and an
analysis of the remaining data can support a field dependent
$\gamma_{\text{eff}}$ \cite{CommentTakahashi}. Torque measurements
performed on larger crystals were reported by Zehetmayer {\em et
al.} \cite{Zehetmayer02}. In this case, data spanning the range of
temperatures from  $5$ to $30\,{\mathrm{K}}$ and measured in
fields of $5$, $10$, and $20\,{\mathrm{kOe}}$ were collected. From
the report it is not clear which field values were used at what
temperatures, and what was the extent of irreversibility, but the
values reported are all in the range between $3.5$ and $4.5$,
monotonously decreasing with increasing temperature. Finally,
Perkins {\em et al.} \cite{Perkins02} report one measurement of
the transverse magnetization with a vibrating sample magnetometer.
At $25\,{\mathrm{K}}$ and $10\,{\mathrm{kOe}}$, they obtained
$\gamma_{\mathrm{eff}}\thickapprox 2.1$.

The different values reported are not completely inconsistent, if
the effective anisotropy $\gamma_{\mathrm{eff}}$ rises with $H$ in
low fields, but eventually saturates to values approaching the
upper critical field anisotropy $\gamma_H$ in fields of the order
of $10\,{\mathrm{kOe}}$. For a thorough analysis, it would be
necessary to have for each measurement the exact information about
field and temperature condition (to assess the limitation $H \ll
H_{\mathrm{c2}}$) as well as the extent of irreversibility (given
that irreversibility leads to an overestimation of
$\gamma_{\mathrm{eff}}$, as discussed above).

However, given the difference between the penetration depth and
coherence length anisotropies \cite{note1} suggested by
theoretical calculations \cite{Kogan02,Miranovic03}, it is
instructive to check the meaning of $\gamma_{\text{eff}}$ and the
anisotropies appearing in Eq.\ (\ref{tau_rev}) in detail. This is
underlined by the difference between $\gamma_{\mathrm{eff}}$ and
the upper critical field anisotropy $\gamma_H$. The anisotropy
parameter enters Eq.\ (\ref{tau_rev}) twice via $\epsilon
(\theta)$. In a first approximation
\cite{Karpinski03SST,AngstPhD}, the appearance outside of the
logarithm (determined by the interaction between vortices) can be
thought of as due to the penetration depth anisotropy
$\gamma_\lambda$, while the appearance in the logarithm (involving
the vortex core cutoff) is linked to the coherence length
anisotropy $\gamma_H$ (see also Ref.\ \cite{Tinkham_intro}). A
field dependence of the effective anisotropy
$\gamma_{\mathrm{eff}}$ similar as shown in Fig.\ \ref{gammaeffH}
follows directly from a penetration depth anisotropy much smaller
than the upper critical field (or coherence length) anisotropy,
even if they are both field independent
\cite{Karpinski03SST,AngstPhD}. Recently, the torque in the London
limit has been rederived by Kogan \cite{Kogan02b} for the general
case of $\gamma_ \lambda \neq \gamma_H$. The generalized formula
is \cite{Kogan02b}:
\begin{equation}
\tau = \frac{V \Phi_{\circ}H}{64 \pi^2 \lambda_{ab}^2} \left ( 1 -
\frac{1}{\gamma_{\lambda}^2} \right ) \frac{\sin 2
\theta}{\epsilon_{\lambda}(\theta)} \left [ \ln \left (
{\frac{\eta H_{\text{c2}}^{\|c}}{H}} \frac{4 e ^2
\epsilon_{\lambda}(\theta)}{(\epsilon_{\lambda}(\theta) +
\epsilon_H (\theta))^2} \right ) - \frac{2
\epsilon_{\lambda}(\theta)}{\epsilon_{\lambda}(\theta) +
\epsilon_H (\theta)} \left ( 1 +
\frac{\epsilon_{\lambda}'(\theta)}{\epsilon_{\lambda}'(\theta)}
\right ) \right ], \label{tau_new}
\end{equation}
where $\epsilon_{\lambda , H}(\theta)=( \cos^2 \theta + \sin^2
\theta / \gamma_{\lambda , H}^2 )^{1/2}$ \cite{note_epsilon},
$\gamma_\lambda = \lambda_c / \lambda_{ab}$, $\gamma_H = \xi_{ab}
/ \xi_c$, $(\ldots)'$ denotes differentiation with respect to the
angle $\theta$, $V$ is the sample volume, and $e = 2.718\ldots$

Based on Eq.\ (\ref{tau_new}) and the calculations of Kogan and
Miranovic \cite{Kogan02,Miranovic03}, Kogan predicted a sign
reversal of the torque at low temperatures in MgB$_2$. We measured
additional torque vs angle curves at low temperatures and in
relatively low fields to experimentally check this prediction.

\begin{figure}[tb]
\centering
\includegraphics[width=0.8\linewidth]{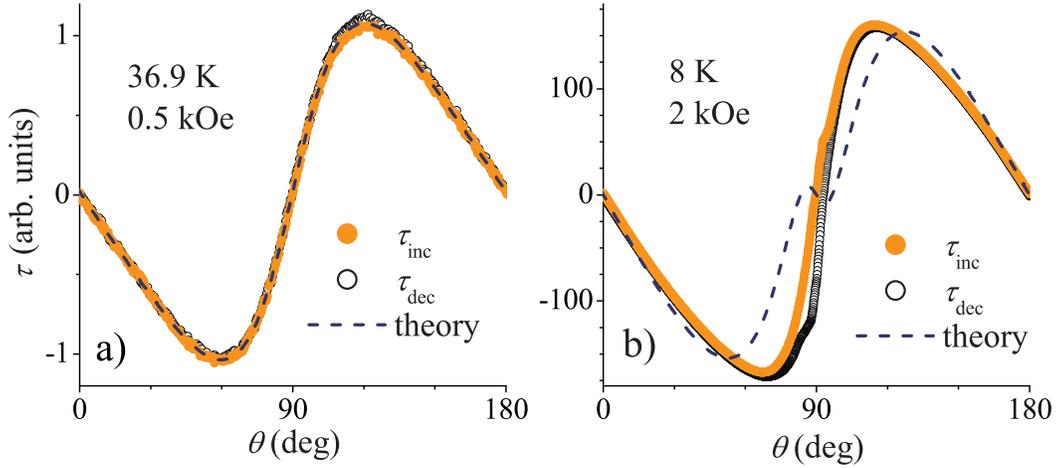}
\caption{ a) Torque $\tau$ vs angle $\theta$ measured close to
$T_c$. Dashed line: $\tau(\theta)$ calculated by Kogan
\cite{Kogan02b} with $\gamma _\lambda = \gamma _H = 2$. b) $\tau$
vs $\theta$ at low temperature. Dashed line: $\tau(\theta)$
calculated \cite{Kogan02b} with $\gamma _\lambda = 2$, $\gamma _H
= 6$. After Ref.\ \cite{Angst03Rio}.} \label{taurevlowT}
\end{figure}

Figure \ref{taurevlowT}a) shows a $\tau(\theta)$ curve measured
(on a different crystal, crystal D) in the mixed state close to
$T_c \simeq 38.5\,{\mathrm{K}}$. Near $T_c$, the difference
between $\gamma _\lambda$ and $\gamma _H$ is small, in agreement
with theoretical predictions \cite{Miranovic03,Kogan02b}. The
common anisotropy value $\gamma = 2$ also is in good agreement
with the data collected on crystal A (see Fig.\ \ref{gammaeffH}).
The $\tau(\theta)$ curve measured at low $T$ [Fig.\
\ref{taurevlowT}b)] has the same sign as the one measured close to
$T_c$, i.e., there is no sign change as expected \cite{Kogan02b}
for $\gamma _\lambda \ll \gamma _H$. For $\gamma _\lambda$
moderately lower than $\gamma _H$, Kogan \cite{Kogan02b} predicted
a sign change only in an angular region close to
$90\,{\mathrm{deg}}$, illustrated with a dashed line in Fig.\
\ref{taurevlowT}b). Such a partial sign change is also not
observed, the maximum angular region where it could occur given by
the irreversible region (the slight asymmetry in the
irreversibility is due to thickness variations of the crystal).
Comparing the data with curves calculated according to Kogan
\cite{Kogan02b}, with $H_{c2}^{\|c}=30\,{\mathrm{kOe}}$,
$\gamma_H=6$ (see Sec.\ \ref{Hc2}) and various $\gamma _\lambda$,
we conclude that $\gamma _\lambda$ has to be at least $2.6$,
considerably higher than currently available theoretical estimates
\cite{Kogan02,Golubov02b}. Alternatively, if $\gamma _H$ in $0.2\,
{\mathrm{T}}$ is much smaller than in $H \approx H_{c2}$
\cite{note1}, the absence of a sign reversal is compatible with
smaller $\gamma _\lambda$. However, we should mention that the
best description of the data with Eq,\ (\ref{tau_new}) is given by
$\gamma _\lambda \approx \gamma _H \approx 3.3$. This value is
also included in the summary of Fig.\ \ref{gammaeffH}
($\blacktriangledown$). This low temperature value is higher than
the values obtained on crystal A in the same field close to
$T_{\mathrm{c}}$, indicating that the anisotropy in fixed field is
no longer temperature independent over a larger $T$ interval.
Concerning the experimental check of the predictions based on Eq.\
(\ref{tau_new}), we note that no total or partial sign reversal
was found in $\tau (\theta)$ curves measured on various crystals
at low temperature in fields between $300\,{\mathrm{Oe}}$ and
$15\,{\mathrm{kOe}}$. However, for the curves measured in the
lowest fields, a small partial sign change may be hidden by
irreversibiliy remaining after shaking.

We think that the discrepancy between the torque data (Fig.\
\ref{taurevlowT}) and the predictions by Kogan \cite{Kogan02b} is
due to one or both of the following reasons:
\begin{itemize}
 \item The calculations of Kogan and Golubov \cite{Kogan02,Golubov02b} are
 valid for the limit of very low fields, whereas $\gamma_H$ was
 calculated for the upper critical field, i.e., in very high fields
 (cf.\ Sec.\ \ref{Hc2}). Instead of $\gamma _\lambda$ and the
 coherence length anisotropy $\gamma _H$ being very different and
 field independent, they could very well be similar or equal, both
 rising with increasing field. Furthermore, for the case of
 arbitrary field, it might be more appropriate to directly start
 with a two band model.
 \item Both Eq.\ (\ref{tau_rev}) and Eq.\ (\ref{tau_new}) are
 based on the London model. In a wide region of the mixed state in
 MgB$_2$, the ``no core overlap'' condition of the London model
 seems highly doubtful due to the ``giant vortex cores'' in the
 $\pi$ band \cite{Eskildsen02}.
\end{itemize}

We first discuss the second possible reason. As mentioned already
in Sec.\ \ref{Hc2}, a hypothetical $\pi$ bands superconductor with
no coupling to the $\sigma$ bands but the same superconducting gap
$\Delta_\pi$ would have a very low upper critical field:
$H_{\mathrm{c2}}^{\pi}(0) \sim 3\,{\mathrm{kOe}}$. This follows
directly from the low superconducting gap in the $\pi$ bands,
using Eqs.\ (\ref{xigap}) and (\ref{xiHc2}). Due to the near
isotropic nature of the $\pi$ bands, no significant anisotropy is
expected for $H_{\mathrm{c2}}^{\pi}$. To the small gap and low
isotropic upper critical field corresponds a large
(direction-independent) coherence length $\xi_\pi (0) \sim
30\,{\mathrm{nm}}$, setting also the radius of the vortex cores.
Hence, for a pure ``$\pi$ bands superconductor'', the limitation
to the London model given by ``no core overlap'' or $H \ll
H_{\mathrm{c2}}$ translates (for all field directions) to $H \ll
3\,{\mathrm{kOe}}$.

The coupling between $\pi$ and $\sigma$ bands by interband
pairing, quasiparticle transfer etc.\ transforms
$H_{\mathrm{c2}}^{\pi}$ from an upper critical field to a mere
crossover in the middle of the mixed state of MgB$_2$. However,
even as a broad crossover, $H_{\mathrm{c2}}^{\pi}$ may be expected
to set a new field scale significantly affecting the mixed state
properties. Furthermore, the whole vortex structure should be
affected, with each vortex having two cores, one in the $\sigma$
bands and another one with much larger radius in the $\pi$ bands.
For both points there is considerable experimental support.

The two superconducting gaps are for example resolved in point
contact spectroscopy measurements, as mentioned in Sec.\
\ref{twoband}. The feature associated with the smaller ($\pi$) gap
was shown to disappear by application of a magnetic field of the
order of $10\,{\mathrm{kOe}}$, independent of the field direction.
The $H_{\mathrm{c2}}^{\pi}$ crossover is also evidenced in a
abrupt change of the field dependence of the specific heat
coefficient $\tilde{\gamma}(H)\equiv \lim_{T \rightarrow 0}
C(H)/T$ (where $C$ is the specific heat) on single crystals
\cite{Bouquet02}. The coefficient $\tilde{\gamma}$ is a (bulk)
probe of the quasiparticle density of states. As far as vortex
core overlap is negligible, each vortex is creating the same
number of quasiparticles in its core, leading to
$\tilde{\gamma}(H)=\tilde{\gamma}_{\mathrm{n}}H/H_{\mathrm{c2}}$,
where $\tilde{\gamma}_{\mathrm{n}}$ is the normal state specific
heat coefficient. The unusual $\tilde{\gamma}(H)$ dependence in
MgB$_2$ can be well explained by an isotropic
$H_{\mathrm{c2}}^{\pi}\sim 3-4\,{\mathrm{kOe}}$, above which the
vortex cores in the $\pi$ band overlap so strongly that the
contribution of the $\pi$ bands to $\tilde{\gamma}$ reaches its
normal state value $\tilde{\gamma}_{\mathrm{n}}^{\pi}$
\cite{Bouquet02} (see also Nakai {\em et al.} \cite{Nakai02}).
Similar arguments apply for the field dependence of the thermal
conductivity, which is a probe of delocalized quasiparticles
\cite{Sologubenko02}. These arguments are strongly supported by
the direct observation of large vortex cores in the $\pi$ bands
with scanning tunnelling spectroscopy (STS) \cite{Eskildsen02}. In
these measurements, huge core radii of about $50\,{\mathrm{nm}}$
are observed, which using Eq.\ (\ref{xiHc2}) would correspond to
$H_{\mathrm{c2}}^{\pi}\sim 1.3\,{\mathrm{kOe}}$. Consequently, a
very significant core overlap is found already in fields of
$2\,{\mathrm{kOe}}$. On the other hand, the vortex lattice in the
$\pi$ bands was still visible in fields as high as
$5\,{\mathrm{kOe}}$, which proves that superconductivity remains
in the $\pi$ bands due to coupling with the $\sigma$ bands even
above $H_{\mathrm{c2}}^{\pi}$. This means that the order parameter
in the $\pi$ bands is not simply gone, but still spatially
varying. The difference in the values of $H_{\mathrm{c2}}^{\pi}$
as deduced from different experiments is simply a consequence of
its nature as a crossover.

The mixed state of a two band superconductor was recently
considered theoretically by Koshelev and Golubov
\cite{Koshelev03}, based on the same approach as the upper
critical field calculations of Refs.\ \cite{Gurevich03,Golubov03}
and neglecting variations of the magnetic field. Koshelev and
Golubov \cite{Koshelev03} pointed out that the behavior of the
weak band ($\pi$ bands) partial density of states (as seen by STS
and relevant for the field dependent specific heat and thermal
conductivity) can be rather different from the behavior of the
weak band order parameter. On the one hand, the crossover in the
(spatially maximal) order parameter field dependence below and
above the weak band ``upper critical field'' is much smoother than
the one of the (spatially averaged) density of states. On the
other hand, the length scale for the variation of the weak band
order parameter is closer to the strong band ($\sigma$ bands)
length scale than would be expected from the length scale of the
variation of the weak band partial density of states (i.e., from
the large $\pi$ band vortex cores observed by STS). With regard to
the applicability of the London model, the order parameter is the
more relevant quantity, since it is the order parameter that is
coupled to the supercurrent density, and thus to the local field
variation around the vortex center and to the interaction between
vortices \cite{Tinkham_intro}. From the relatively smooth
variation of the maximum order parameter in the weak band we
conclude that there is still a large spatial $\pi$ bands order
parameter variation up to the bulk upper critical field, which
points against the applicability of the London model in fields
larger than $\sim H_{\mathrm{c2}}^{\pi}$. We mention that recent
preliminary torque measurements also point against the London
model being appropriate in MgB$_2$ for $H >
H_{\mathrm{c2}}^{\pi}$.

Before considering the other possible reason for the discrepancy
between torque results and the predictions of Kogan
\cite{Kogan02b}, it is useful to list anisotropy effects in the
mixed state observed by other methods. The anisotropy of the
thermal coefficient of the specific heat $\tilde{\gamma}(H)$ (see
above) measured for $H\|c$ and $H\|ab$ can be formally defined as
the ratio of fields applied perpendicular and parallel to the $c$
axis that yield the same value of $\tilde{\gamma}$
\cite{Bouquet02}. Normally, this gives a constant value
$\gamma_{c_p}$ identical to the upper critical field anisotropy
(this follows directly from
$\tilde{\gamma}(H)=\tilde{\gamma}_{\mathrm{n}}H/H_{\mathrm{c2}}$).
In MgB$_2$, however, Bouquet {\em et al.} \cite{Bouquet02} found
$\gamma_{c_p}$ to be field dependent: It has a value of
$\gamma_{c_p}\simeq 1$ for $H\lesssim 10\,{\mathrm{kOe}}$ and then
rises roughly linearly up to the $H_{\mathrm{c2}}$ anisotropy when
the field approaches the upper critical field. This dependence is
reminiscent of the field dependence of $\gamma_{\mathrm{eff}}$
discussed earlier, but is the rise of $\tilde{\gamma}$ starting
only in elevated fields is an essential difference. The advantage
of these specific heat experiments was the sensitivity only to the
vortex cores, without the added complication of another length
scale (the penetration depth). Indeed, Bouquet {\em et al.} were
able to separate the contributions of $\sigma$ and $\pi$ bands.
The contribution from the $\sigma$ band has a constant anisotropy
of about $6$ and the contribution from the $\pi$ band is
isotropic, both consistent with the calculated electronic
anisotropy in these bands (see Sec.\ \ref{twoband}). The effective
anisotropy $\gamma_{c_p}$ is just a weighted average. The field
dependence of $\gamma_{c_p}$, sensitive to the vortex cores only,
does not indicate a field dependence of the coherence length, but
rather that phenomenological formulas should take into account
that there are two different coherence lengths in MgB$_2$,
$\xi_{\pi}$ and $\xi_{\sigma}$. Whenever this is not taken into
account, the anisotropy of the best ``effective coherence length''
needed to describe experimental data {\em has} to be field
dependent due to the different ``intrinsic upper critical fields''
of the $\sigma$ and $\pi$ bands.

An important anisotropy effect is the distortion of the vortex
lattice in fields tilted away from the $c$ axis, observed for
example in the strongly anisotropic cuprate superconductors by
small angle neutron scattering (SANS) \cite{Yethiraj93}: the
nearest neighbors fall no longer on a circle, but on an ellipse
with axis ratio $r$ given by $(\cos^2 \theta + \sin^2 \theta /
\gamma^2)^2$ in superconductors with a single constant anisotropy
parameter. In superconductors with no single anisotropy, we can
define a lattice distortion anisotropy $\gamma_{\mathrm{VL}}$:
\begin{equation}
r = \left ( \cos^2 \theta + \sin^2 \theta / \gamma_{\mathrm{VL}}
^2 \right ) ^2 \label{distortion}
\end{equation}
For $\gamma_\lambda \neq \gamma_\xi$ the distortion is determined
by the penetration depth anisotropy in the London limit
\cite{Campbell88,Kogan02b}:
$\gamma_{\mathrm{VL}}=\gamma_{\lambda}$. Recently, SANS
measurements were carried out on MgB$_2$ powder samples in
$5\,{\mathrm{kOe}}$ at $2\,{\mathrm{K}}$ by Cubitt {\em et al.}
\cite{Cubitt03}. In randomly aligned powder, each grain carries a
vortex lattice of its own, leading to a ring of diffraction
instead of resolved diffraction spots. The random orientation of
the grains and the vortex lattice distortion together lead to a
broadening of this ring of diffraction. Comparing the measured
scattered intensity distribution to the theoretically expected
distribution, Cubitt {\em et al.} deduced an upper limit of the
vortex lattice anisotropy under the given conditions:
$\gamma_{\mathrm{VL}} < 1.32$. In case that the London model is
applicable in MgB$_2$ in $5\,{\mathrm{kOe}}$ at $2\,{\mathrm{K}}$,
this would then correspond to the upper limit of the penetration
depth anisotropy. In light of the discussion above, the
applicability of the London model for the experimental conditions
seems questionable. Indeed, according to the calculations of Dahm
and Schopohl \cite{Dahm02}, the vortex lattice distortion
anisotropy in fields close to the upper critical field should be
similar to $\gamma_H$. The method used by Cubitt {\em et al.}
\cite{Cubitt03} furthermore depends for example on model
assumptions of the grain alignment distribution under the
influence of the magnetic field and on the validity of the angular
scaling of Eq.\ (\ref{distortion}) [see also Sec.\ \ref{Hc2}].
Later, SANS was also used to determine the vortex lattice
structure and distortion on a single crystal \cite{Eskildsen_pc}.
Measurements with $H\|c$ indicate a structural transformation of
the vortex lattice in fields between about $5$ and
$9\,{\mathrm{kOe}}$, which is likely associated with the
suppression of superconductivity in the $\pi$ bands above
$H_{\mathrm{c2}}^{\pi}$. Measurements with $H$ tilted by
$45\,{\mathrm{deg}}$ away of the $c$ axis show a vortex lattice
anisotropy rising superlinearly with the field from
$\gamma_{\mathrm{VL}}\simeq 1.3$ in $2\,{\mathrm{kOe}}$ to roughly
$3.5$ in $5\,{\mathrm{kOe}}$. Differences between the value
obtained in $2\,{\mathrm{kOe}}$ and the value obtained from
scanning tunnelling spectroscopy in the same field, but for
$H\|ab$ \cite{Eskildsen_pc} may indicate a deviation of the
angular dependence from Eq.\ (\ref{distortion}). Similar
deviations are discussed in Sec.\ \ref{Hc2}. A deviation from Eq.\
(\ref{distortion}) could also explain the discrepancy to the SANS
measurement on a powder sample \cite{Cubitt03}. The identification
of the observed $\gamma_{\mathrm{VL}}$ with the penetration depth
anisotropy is still questionable. However, the extrapolated low
field lattice anisotropy $\lim_{H \rightarrow 0}
\gamma_{\mathrm{VL}}(H) \approx 1.25$ should be equal to the low
field penetration depth anisotropy, since in low enough fields
neither the vortex cores in the $\sigma$ bands nor the ones in the
$\pi$ bands overlap and the London model should apply. Indeed, the
value of $\sim 1.25$ is consistent with the low field theoretical
calculations \cite{Kogan02,Golubov02b}.

For further quantitative analyses it will be necessary to make
theoretical progress beyond the simple (one band) London model,
building a sophisticated model of the mixed state built by
vortices with two cores (or a ``double core'') as well as two
sorts of superconducting carriers. Steps into this direction were
undertaken \cite{Nakai02,Babaev02,Gurevich03b,Koshelev03}, but
what is lacking so far is a model of the variation of the local
field around such ``double core'' vortices, particularly for the
case of a large overlap of the larger of the cores. This case is
very important, since it applies for the largest region of the
mixed state of MgB$_2$ ($H \gtrsim H_{\mathrm{c2}}^{\pi}$, see
Fig.\ \ref{gammaHvsT}).

Qualitatively, the picture emerging seems to be more clear. As in
the analysis of zero field specific heat and penetration depth
data \cite{Bouquet01c}, it is important to consider the
contributions of both sets of bands on any mixed state property.
Each of the two sets of bands has its own set of length scales,
i.e., coherence length $\xi$ and penetration depth $\lambda$, the
latter of which is linked to the partial superfluid density [see
Eq.\ (\ref{lambdadef})]. In the case of the $\pi$ bands these
length scales are isotropic, whereas in the case of the $\sigma$
bands, they are highly anisotropic, with an anisotropy of about
$6-7$, in accordance to the band structure calculations (see Sec.\
\ref{twoband}). These intrinsic anisotropies of the $\pi$ and the
$\sigma$ bands probably do not depend on field or temperature. Any
mixed state property is then determined by the relative
contributions from both bands and if described in terms of
traditional single band models has to be assigned some effective
anisotropy, depending on the contributions of the two sets of
bands. At low temperatures and fields, both bands contribute
roughly equally to the properties, in accordance to their similar
contribution to the density of states at the Fermi level (see
Sec.\ \ref{twoband}). Since the superconductivity is weaker in the
$\pi$ bands (smaller gap), it gets depressed faster by
perturbations such as an external field or thermal excitations.
However, due to the coupling with the $\sigma$ bands, it remains
superconducting until the $\sigma$ bands become also
nonsuperconducting: the whole MgB$_2$ is either in the
superconducting or in the normal state. Upon application of an
external magnetic field, superconductivity is depressed much
faster in the weak $\pi$ bands, implying a decrease of the
relative $\pi$ bands contribution to any mixed state property.
Consequently any effective anisotropy used to describe the
property has to rise with increasing field. How exactly that rise
occurs depends on the quantity in question and the definition of
the effective anisotropy. This applies for angle dependent torque
measurements as well as for specific heat and vortex lattice
distortion, as we have seen above. Increasing the temperature
instead of the field in principle has a similar effect. However,
additionally, thermally excited quasi-particles lead to an
increased mixing of the two sets of bands, until, extremely close
to $T_{\mathrm{c}}$, the two band character of MgB$_2$ is no
longer important for the mixed state properties and conventional
anisotropic Ginzburg-Landau theory (AGLT) applies (cf.\ Sec.\
\ref{Hc2}). Below this tiny region, AGLT is no longer appropriate
to describe the mixed state of MgB$_2$, and even phenomenological
theories (Ginzburg-Landau \cite{Babaev02}, London, $\ldots$)
should be developed and used in a two band form.

\section{Peak effect and vortex matter phase diagram}
\label{PE}

In Sec.\ \ref{mixedstate}, we examined anisotropy effects in the
mixed state, which is bounded by the upper and lower critical
fields (see Sec.\ \ref{Hc2}). From the study of cuprate
superconductors is known that the mixed state phase diagram
contains more transition lines than the upper and lower critical
fields. Identified were for example a melting transition between a
quasi-ordered vortex lattice, called Bragg glass, and a disordered
vortex fluid \cite{Schilling96}, as well as an order-disorder
transition between the Bragg glass and a highly disordered, glassy
phase \cite{Giamarchi94,Khaykovich96,Giller97}. The latter
transition \cite{note_Avraham} has been observed also in low $T_c$
superconductors, such as NbSe$_2$ \cite{Ravikumar01,Marchevsky01},
and even in the elemental superconductor Nb \cite{Ling01}, but not
in ultra-pure Nb crystals \cite{Forgan02}. This transition is
generally associated with a peak in the critical current density
$j_c$ and pronounced history effects.

MgB$_2$ is, particularly concerning the importance of thermal
fluctuations and the value of $\kappa=\lambda/\xi$, intermediate
between the high $T_{\mathrm{c}}$ cuprates and low
$T_{\mathrm{c}}$ superconductors. Studying the vortex matter phase
diagram of MgB$_2$ may thus help in understanding the phase
diagrams of various superconductors in a unified way.

\begin{figure}[tb]
\centering
\includegraphics[width=0.85\linewidth]{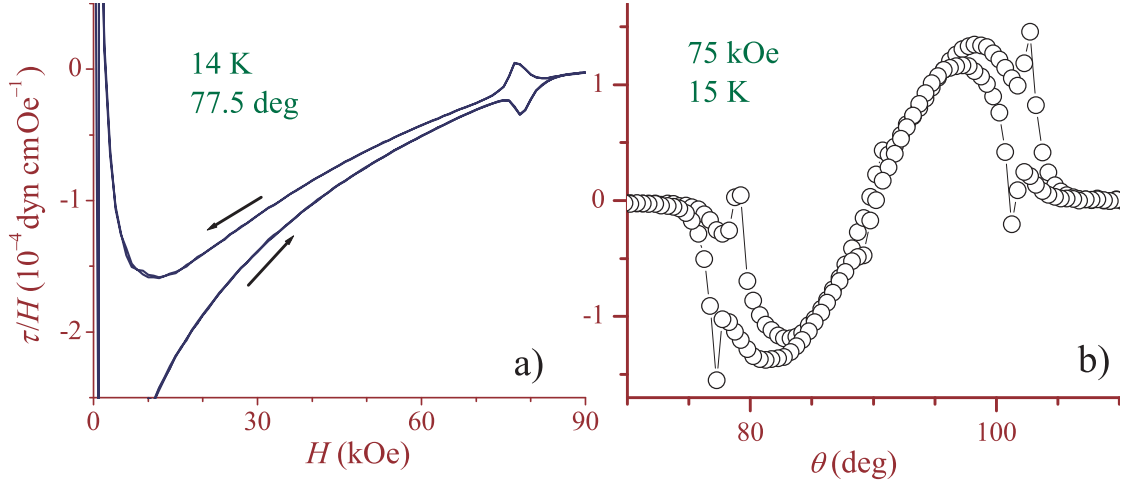}
\caption{ Peak effect in the irreversible torque of MgB$_2$ single
crystal B: a) Torque $\tau/H$ vs field $H$ measured at
$14\,{\text{K}}$ and $\theta=77.5\,{\mathrm{deg}}$. The direction
of the field change is indicated by arrows. b) $\tau/H$ vs angle
$\theta$ in $75\,{\mathrm{kOe}}$ at $15\,{\mathrm{K}}$. After
Refs.\ \cite{MgB2PE,Angst03MgB2PhyC}.} \label{PE1}
\end{figure}

In single crystals of MgB$_2$, a quasi-ordered vortex structure
has been observed in low fields by scanning tunnelling
spectroscopy \cite{Eskildsen02} and more recently clear Bragg
peaks were observed by small angle neutron scattering
\cite{Eskildsen_pc}, showing that at least under some conditions a
Bragg glass is the stable vortex phase. Since, by tuning the
amount of quenched random point-like disorder, the stabilization
of a highly disordered phase can always be favored, an
order-disorder transition in fields below $H_{c2}$ should be
observable in MgB$_2$ as well, at least for certain impurity
concentrations. We therefore looked for signatures of this
transition by measuring torque as a function of field and angle.

One of the torque vs field curves measured on crystal B is shown
in Fig.\ \ref{PE1}a). For better comparison with magnetization
curves, $\tau/H$ vs $H$ is shown. A pronounced and sharp peak
effect (PE) in fields close to, but clearly distinct from,
$H_{\text{c2}}$, can be seen. The peak effect was observed in
$\tau$ vs $\theta$ measurements in fixed field as well, an example
is shown in Fig.\ \ref{PE1}b). However, several features of the PE
are better manifest in $\tau$ vs $H$ measurements at fixed angle,
on which we concentrate below.

\begin{figure}[tb]
\centering
\includegraphics[width=0.85\linewidth]{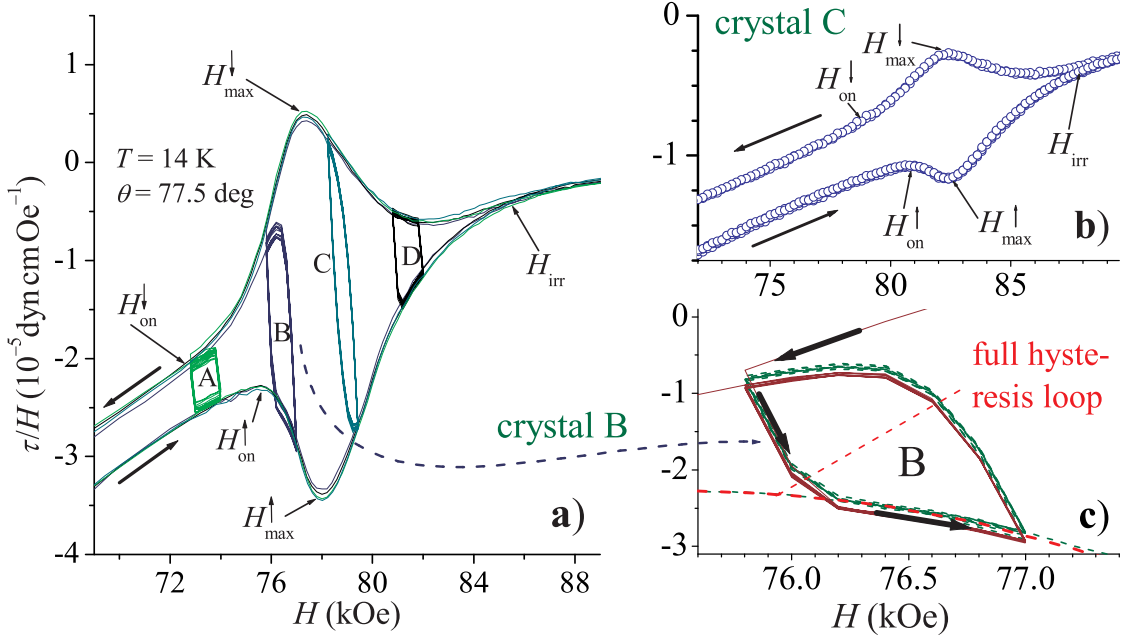}
\caption{ a) Magnification of the peak effect region of the curve
shown in Fig.\ \ref{PE1}a) for crystal B. Also shown are some of
the minor hysteresis loops MHL (see text) measured, labelled
A$-$D. b) peak effect region of a $\tau/H$ vs.\ $H$ curve measured
on crystal C, under the same conditions as were used for panel a).
c) Magnification of MHL B shown in panel a). MHL started from the
field increasing branch of the full hysteresis loop are shown as
dotted lines (also indicated by thick arrows), whereas those
started from the field decreasing branch are shown as full lines.
The dashed line shows the field increasing branch of the (normal)
full hysteresis loop. After Refs.\ \cite{MgB2PE,Angst03MgB2PhyC}.}
\label{MHL}
\end{figure}

A magnification of the PE region of the $\tau(H)$ curve shown in
Fig.\ \ref{PE1}a) is presented in Fig.\ \ref{MHL}a). The peak is
well pronounced and very sharp (full width at half maximum $\sim
4\,{\text{kOe}}$ in comparison to the location in fields of about
$78\,{\text{kOe}}$). Various characteristic fields are indicated:
The maximum of the peak for field increasing
($H_{\text{max}}^{\uparrow}$) and decreasing
($H_{\text{max}}^{\downarrow}$) branch of the hysteresis loop, and
the onsets of the peak, $H_{\text{on}}^{\uparrow}$ and
$H_{\text{on}}^{\downarrow}$. The separation of the two onset
fields is larger than the separation of the maximum fields,
similar to the case of the cuprate superconductors
\cite{note_Sr124}. Also indicated is the irreversibility field
$H_{\text{irr}}$, where the two branches of the hysteresis loops
meet. The peak resembles qualitatively the peaks observed in
NbSe$_2$ \cite{Eremenko02,Ravikumar01} and CeRu$_2$ \cite{Roy00}.
In NbSe$_2$, there are strong indications that the peak effect
indeed is a signature of the order-disorder phase transition
\cite{Marchevsky01}.

Figure \ref{MHL}b) shows a measurement on crystal C with the same
external conditions as in panel a). In this crystal, the PE is
less pronounced, but still clearly discernible. It is located in
slightly higher fields, $H_{\text{max}}^{\uparrow}\simeq
82.4\,{\text{kOe}}$. Crystal C was grown with the same technique
as crystal B, but under slightly different conditions. These two
crystals have pronouncedly different pinning properties,
particularly for field directions close to parallel to the layers
(see Angst \cite{AngstPhD}). The presence of the PE in two
crystals with such pronounced differences strongly indicates that
the PE, or rather its underlying mechanism, is an intrinsic
feature of MgB$_2$. That this is indeed the case is also supported
by the recent observations of similar peak effects on single
crystalline MgB$_2$ from other sources by electrical transport
\cite{Lyard02,Welp03}, ac susceptibility with a Hall probe
\cite{Lyard02,Pissas02}, and torque magnetometry \cite{Cooper_pc}.

The peak effect shown in Fig.\ \ref{MHL} is similar to the peak
effect in NbSe$_2$, where it is attributed to the order-disorder
transition, as we mentioned above. The order-disorder transition
region is accompanied by history effects in the critical current
density (a brief discussion of this point can be found in Ref.\
\cite{AngstPhD}), a feature observed experimentally for example in
NbSe$_2$ \cite{Ravikumar01,Marchevsky01}, and very useful to
identify the transition.

\begin{figure}[tb]
\centering
\includegraphics[width=0.9\linewidth]{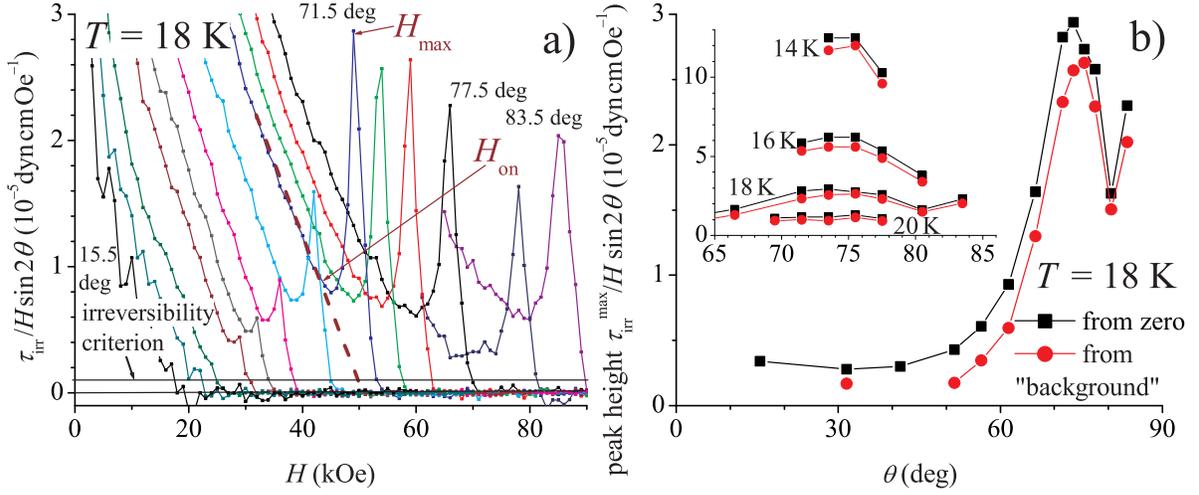}
\caption{a) Variation of the irreversible torque, scaled by $H
\sin 2 \theta$, vs field, at $18\,{\mathrm{K}}$ (crystal B, after
Ref.\ \cite{Angst03MgB2PhyC}). The curves were measured at angles
(from left to right) $\theta=15.5$, $31.5$, $41.5$, $51.5$,
$56.5$, $61.5$, $66.5$, $71.5$, $73.5$, $75.5$, $77.5$, $80.5$ and
$83.5\, {\mathrm{deg}}$. Horizontal lines indicate zero and the
criterion chosen for the determination of the irreversibility line
\cite{note_Hirr_crit}. For the curve measured at $71.5\,
{\text{deg}}$, peak maximum and onset are indicated (see text). b)
Height of the peaks visible in a) vs angle. The peak heights both
from zero, and from the extrapolated background signal without
peak are shown. Inset: Comparison of the angle dependence of the
peak height at different temperatures.} \label{peakheight}
\end{figure}

To investigate possible history dependences of the critical
current density, we performed several minor hysteresis loop (MHL)
measurements in and around the peak of crystal B \cite{MgB2PE}:
The field is cycled up and down by a small amount
($1.2\,{\text{kOe}}$ in steps of $200\,{\text{Oe}}$) several
times, ideally until the loops retrace each other, indicating that
the vortex system reached a stable pinned state in the given field
\cite{Ravikumar01,Roy00}. MHL measured, within full loops, in four
different regions of the PE are indicated in Fig.\ \ref{MHL}a)
(A-D). MHL B is magnified in Fig.\ \ref{MHL}c): The initial
$H^{\uparrow}$ branch of the MHL started from the $H^{\downarrow}$
branch of the full hysteresis loop (FHL) (full line indicated by
arrows) clearly is below the $H^{\uparrow}$ branch of the FHL
(thick dashed), indicating larger hysteresis. This behavior
contradicts Bean's critical state model \cite{Bean62}, where the
hysteresis of partial hysteresis loops can never be higher than
the one of the full loop. It can be explained by the vortex
configuration on the $H^{\downarrow}$ branch of the FHL (where the
MHL was started) having a higher $j_c$ than the vortex
configuration on the $H^{\uparrow}$ branch. Repeated cycling
studies furthermore indicate the meta-stable nature of the this
``high field vortex configuration'', at $76.5\,{\mathrm{kOe}}$
\cite{MgB2PE,AngstPhD}. In MHL A, similar but less pronounced
effects were observed, whereas no history effects were discerned
in studying MHL B, C, and additional MHL far below the PE.

The findings of the detail study of minor hysteresis loops of the
peak effect in MgB$_2$ \cite{MgB2PE,AngstPhD} can be summarized as
follows: Between $H_{\text{on}}^{\downarrow}$ and
$H_{\text{max}}^{\uparrow}$, pronounced history effects occur.
They can be accounted for by the coexistence of a metastable
high-field vortex configuration with high pinning and a stable
low-field, low pinning configuration. Above
$H_{\text{max}}^{\uparrow}$ and below
$H_{\text{on}}^{\downarrow}$, no significant history effects are
observed, indicating that there is only one vortex configuration,
which is stable.  The larger hysteresis width of MHL started from
$H^{\downarrow}$ indicates pinning in the configuration stable
above $H_{\text{max}}^{\uparrow}$ to be stronger than pinning in
the configuration stable below $H_{\text{on}}^{\downarrow}$. This
is exactly what is expected for the order-disorder transition
\cite{Marchevsky01,AngstPhD}. Particularly, the ``overshoot'' of
an MHL over the full hysteresis loop [Fig.\ \ref{MHL}c)] is
difficult to explain without a ``two metastable vortex
configurations'' scenario. Specifically, it cannot be explained by
relaxation effects, since this should always lead to lower
hysteresis of the MHL with respect to full hysteresis loops.

\begin{figure}[tb]
\centering
\includegraphics[width=0.7\linewidth]{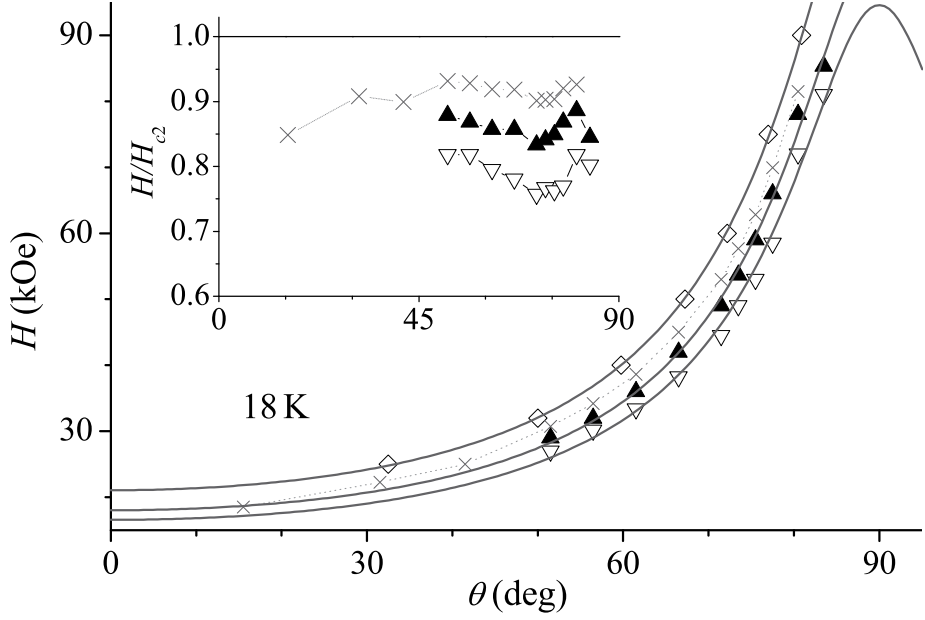}
\caption{ Angle dependence of various characteristic fields at
$18\,{\mathrm{K}}$. Shown are the upper critical field
$H_{\mathrm{c2}}$ ($\diamondsuit$), the irreversibility field
$H_{\mathrm{irr}}$ ($\times$) \cite{note_Hirr_crit}, the peak
maximum field $H_{\max}$ ($\blacktriangle$), and the peak onset
field $H_{\mathrm{on}}$ ($\nabla$). Full lines are fits of the
theoretical $H_{\mathrm{c2}}(\theta)$ dependence [Eq.\
(\ref{Hc2_theta})]. Dashed lines are guides for the eye. Inset:
Angle dependence of reduced [divided by $H_{\mathrm{c2}}(\theta)$,
cf.\ Sec.\ \ref{Hc2}] characteristic fields (using the same
symbols as in the main panel). After Ref.\ \cite{MgB2PE}.}
\label{PEchartheta}
\end{figure}

Figure \ref{peakheight}a) shows the irreversible part $\Delta \tau
(H) = \tau(H^{\downarrow})-\tau(H^{\uparrow})$ of the torque,
scaled by $H \sin 2 \theta$, vs field, at $18\,{\text{K}}$ for
various angles. The scaling was chosen to minimize the angle and
field dependence intrinsic to the torque. Since the peak is not
visible at all temperatures and angles as well as in Fig.\
\ref{MHL}, onsets and maxima were determined from irreversible
torque curves as those shown in Fig.\ \ref{peakheight}a). The
$H_{\text{on}}$ was defined as the field, where the irreversible
torque starts to deviate from a straight line behavior, as
indicated in the figure for the curve measured at
$71.5\,{\text{deg}}$. The onset defined in this way is close to
$H_{\text{on}}^{\downarrow}$ as indicated in Fig.\ \ref{MHL}a),
but we note that with the determination of onsets and maxima from
the irreversible torque, the fine details of the differences in
the field increasing and decreasing branch of the hysteresis loops
are lost. It can be seen in Fig.\ \ref{peakheight}a) that the
height of the peaks varies in a pronounced way with the angle
$\theta$. The peak height is shown in panel b), with and without
subtraction of an extrapolated linear background. It indicates
that the peak effect is most pronounced in the region of angles
between about $70$ and $80\,{\text{deg}}$. This angle region is
independent of temperature, as shown in the inset of Fig.\
\ref{peakheight}b).

\begin{figure}[tb]
\centering
\includegraphics[width=0.8\linewidth]{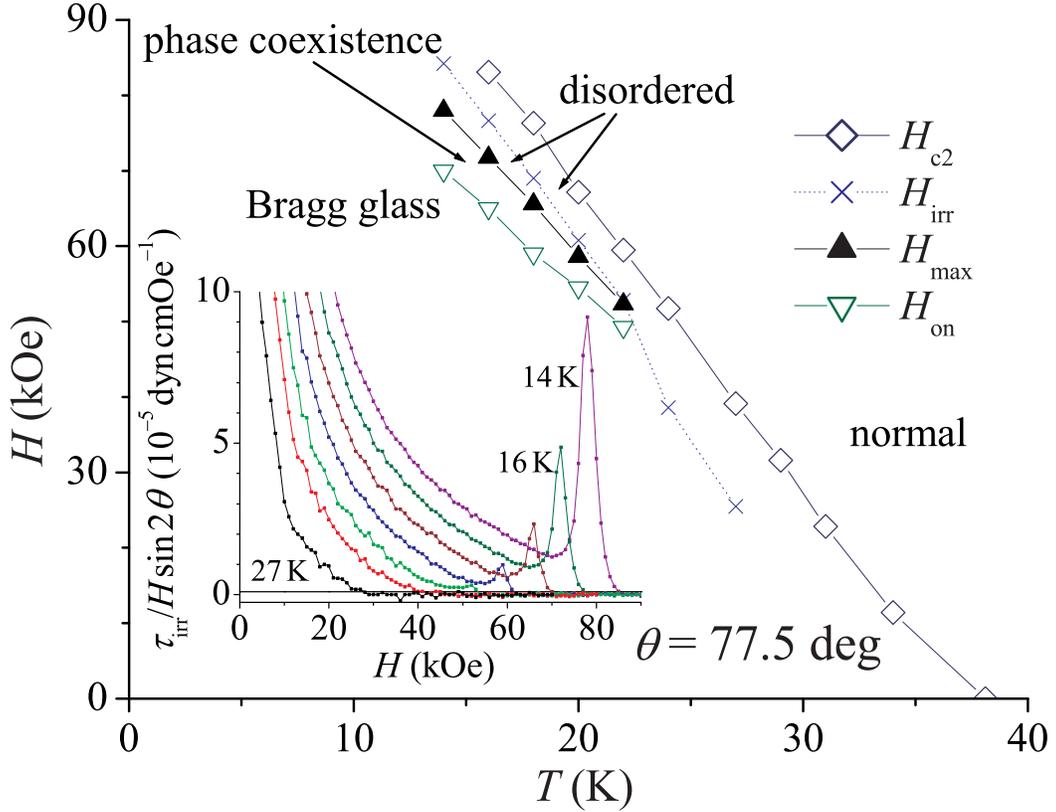}
\caption{Phase diagram of MgB$_2$ single crystal at an angle of
$77.5\,{\mathrm{deg}}$ between the $c$-axis of the crystal and the
applied field, constructed from the curves shown in the inset. The
temperature dependence of the characteristic fields
$H_{\mathrm{c2}}$, $H_{\mathrm{max}}$, and $H_{\mathrm{on}}$ is
given. They mark boundaries between the normal state and the
various phases of vortex matter, indicated in the figure and
discussed in the text. The irreversibility field
$H_{\mathrm{irr}}$ is also shown. After Ref.\ \cite{MgB2PE}.
Inset: Variation of the irreversible torque, scaled by $H \sin 2
\theta$, vs field $H$, at $77.5\,{\mathrm{deg}}$. The curves were
measured at temperatures (from left to right)
$T=27,24,22,20,18,16,$ and $14\,{\mathrm{K}}$. The horizontal line
near zero marks the irreversibility field criterion
\cite{note_Hirr_crit}.} \label{phd775deg}
\end{figure}

One possible explanation for this behavior is an influence of
stacking faults on the peak effect. Although the presence of the
peak effect, and it's location, are not affected by stacking
faults, the extent of hysteresis may be. The difference of how
pronounced the peaks of crystals B and C are [see Figs.\ \ref{PE1}
and \ref{MHL}b)] supports such a scenario. The location in higher
fields of the peak effect in crystal C indicates that there is
less point-like disorder present in this crystal than in crystal
B, if we accept the PE as a signature of the order-disorder
transition \cite{Giamarchi94,Khaykovich96,Giller97}. What is
difficult to explain with only one sort of disorder varying is the
smaller ratio of the hysteresis width within to the hysteresis
width below the PE,
$\tau_{\text{irr}}(H_{\text{max}})/\tau_{\text{irr}}(H_{\text{on}})$,
of crystal C, as compared to crystal B. Both the angular PE
amplitude maximum and the difference in the ratio
$\tau_{\text{irr}}(H_{\text{max}})/\tau_{\text{irr}}(H_{\text{on}})$
of crystals B and C might be attributed to the influence of
non-collective strong pinning by stacking faults in crystal B.
This point was discussed in more detail by Angst \cite{AngstPhD}.

The variation of the peak onsets and maxima with angle at
$18\,{\text{K}}$ is shown, together with $H_{c2}(\theta)$ and
$H_{\text{irr}}(\theta)$ \cite{note_Hirr_crit}, in Fig.\
\ref{PEchartheta}. These characteristic peak fields follow the
angular dependence of $H_{c2}$, as indicated by fits to the
theoretical $H_{c2}(\theta)$ dependence according to the
anisotropic Ginzburg-Landau theory [Eq.\ (\ref{Hc2_theta}) of
Sec.\ \ref{Hc2}], while the angular scaling of the irreversibility
field is less clear. This can be seen also in the inset,
displaying the $\theta$ dependence of the characteristic fields,
reduced by the upper critical field. The onset field is
approximately constant at about $0.8\,H_{c2}$ and the maximum
field at about $0.85\,H_{c2}$. $H_{\text{irr}}$ is located at
about $0.9\,H_{c2}$, but seems to get slightly lower as
$\theta\rightarrow 0$. The good angular scaling of the peak onset
and maximum fields supports the conclusion that they are
unaffected by stacking faults.

Torque measurements are not possible for $H\|c$ or $H\|ab$ and
SQUID measurements performed with $H\|c$ and $H\|ab$ gave no
indication of the peak effect. However, ac susceptibility
measurements on the same crystals indicate the peak effect to be
present both for $H\|c$ and $H\|ab$, confirming that the
underlying mechanism is a feature for all directions of $H$
\cite{Angst03MgB2PhyC,Puzni_prep}. This conclusion is supported
also by recent reports of other authors, in the case of $H\|c$
from transport data \cite{Welp03,Lyard02} and ac susceptibility
with a local Hall probe \cite{Lyard02,Pissas02}, in the case of
$H\|ab$ from transport data \cite{Lyard02}.

The phase diagram for an angle of $77.5\,{\text{deg}}$ between $H$
and the $c$ axis of the crystal (corresponding roughly to the
angle where the PE is most visible), is presented in Fig.\
\ref{phd775deg}. In the inset, scaled irreversible torque vs $H$
curves are shown. The magnitude of the peaks is reduced quickly by
increasing the temperature. The main panel shows the temperature
dependence of the characteristic fields. The abrupt end of
$H_{\text{on}}(T)$ and $H_{\text{max}}(T)$ is due to the decreased
sensitivity of the magnetometer (In fact, the peak effect is
discernible at temperatures up to $\sim 27\,{\mathrm{K}}$ in
measurements of the torque vs angle, such as shown in Fig.\
\ref{PE1}b) \cite{Angst03MgB2PhyC}) and due to the smearing of the
effective pinning landscape by thermal fluctuations
\cite{AngstPhD}, as indicated by the drop of the irreversibility
line.

In a recent report of low frequency ac susceptibility measurements
\cite{Pissas02}, the peak effect was observed for $H\|c$ at
temperatures up to about $25\,{\mathrm{K}}$, and interpreted in
terms of the order-disorder transition as well. Pissas {\em et
al.} \cite{Pissas02} report the transformation of the PE into a
``step-like'' ac susceptibility for the temperature interval
between $25$ and $27.5\,{\mathrm{K}}$, and interpreted this as a
signature of thermal melting. In our case, no step-like feature in
the reversible torque was observed in the continuation of the PE.
It should be emphasized, that thermal melting so far below
$H_{\mathrm{c2}}$ would be at odds \cite{MgB2PE} with theoretical
expectations \cite{Mikitik01}.

The equilibrium order-disorder transition, which corresponds to
$H_{\mathrm{max}}$ \cite{MgB2PE}, is located in fields of about
$0.85\,H_{\mathrm{c2}}$ in crystal B and in about
$0.9\,H_{\mathrm{c2}}$ in crystal C. The peak effect observed in
other crystals by transport was reported to be located even closer
to $H_{\mathrm{c2}}$ \cite{Welp03,Lyard02}. These differences are
natural for a disorder-induced phase transition in crystals with
varying degrees of disorder. Form and location of the PE observed
in MgB$_2$ resembles results obtained on NbSe$_2$ single crystals
with varying degrees of disorder \cite{note_Banerjee99}, but are
rather different from the order-disorder transition in cuprate
superconductors \cite{note_Sr124,AngstPhD}.

%\clearpage

\section{Conclusions}
\label{conc}

We have reviewed superconducting state anisotropy effects in
magnesium diboride MgB$_2$, with the main focus on results
obtained by torque magnetometry. MgB$_2$ is a ``two band'' or
``two gap'' superconductor: the superconductivity is Fermi sheet
dependent in an extreme way, with a large gap (strong
superconductivity) on the $\sigma$ Fermi sheets and a roughly
three times smaller gap on the $\pi$ Fermi sheets. This is
possible because of the very different coupling to the lattice
vibrations in these two sets of bands and because of there is very
low hybridization between them, resulting in a very low impurity
scattering rate between them. Pronounced and unusual anisotropy
effects originate in this two band superconductivity and in the
very different electronic anisotropies of the two sets of bands.

One of the unusual anisotropy effects in the superconducting state
is the very pronounced temperature dependence of the upper
critical field anisotropy $\gamma_H$, falling from a value close
to (with some sample dependence) the electronic anisotropy of the
$\sigma$ bands ($\sim 6.8$) at zero temperature to low values of
the order of about $2$ for temperatures approaching
$T_{\mathrm{c}}\simeq 39\,{\mathrm{K}}$. There is a good
qualitative agreement between various experiments as well as
theoretical studies on this. The origin of this temperature
dependence of $\gamma_H$ is the increased ``mixing'' of the two
sets of bands by thermally excited quasiparticles at higher
temperatures $-$ at low temperatures the upper critical field is
mainly determined by the stronger superconducting $\sigma$ bands,
which has the higher field scale. The interplay between the
different field scales, anisotropies, and interband mixing also
leads to the observed deviations of the angular dependence of
$H_{\mathrm{c2}}$ from conventional behavior. No general consensus
has yet emerged concerning the impact of the different intraband
impurity scattering rate on the upper critical field. Concerning
the lower critical fields, the available experimental data is
sparse, preventing the drawing of firm conclusions.

The anisotropy effects on the upper critical field, as well as on
other quantities, implies a breakdown of the standard (one band)
anisotropic Ginzburg-Landau theory, except in a tiny (less than
$1\,{\mathrm{K}}$) region below $T_{\mathrm{c}}$. Below this
region, microscopic calculations have to be used, or at least a
two band formulation of the anisotropic Ginzburg-Landau theory
involving two order parameters and two sets of length scales $\xi$
and $\pi$, as well as some description of the coupling between
these two components.

In the mixed state, experimentally deduced anisotropies are
different from different quantities, and also often depend on the
model used to analyze the data. However, all experimental results
[at least those spanning an extended field region including (at
low temperatures) the range of $1-15\,{\mathrm{kOe}}$] find
anisotropies that are monotonically increasing as a function of
the magnetic field and that are consistent with reaching
$\gamma_H$ for $H\rightarrow H_{\mathrm{c2}}$ or already in lower
fields. In the case of torque magnetometry, describing the data
with an equation allowing for different anisotropies of the
penetration depth and the coherence length yields anisotropies of
these length scales that are rather similar.

The field dependent effective anisotropies of various mixed state
quantities are qualitatively well described by a crossover from a
low field region, where the quantities are determined by similar
contributions from the isotropic $\pi$ and the anisotropic
$\sigma$ bands, to a high field region, where they are (almost)
determined by the $\sigma$ bands alone. This crossover is due to
the very low ``intrinsic upper critical field of the $\pi$ bands''
$H_{\mathrm{c2}}^{\pi}$, an order of magnitude lower than the one
(in $c$ direction) of the $\sigma$ bands, because of the weaker
superconductivity (smaller gap) in the $\pi$ bands. In decoupled
$\pi$ and $\sigma$ superconductors, there would be a sharp
transition at $H_{\mathrm{c2}}^{\pi}$, but the coupling between
the two sets of band transforms this into a smooth crossover,
which in addition may be qualitatively different for the gaps and
the quasi-particle excitations in the vortex cores. This may also
explain the qualitatively different forms of the field dependence
of the anisotropy as observed for example in torque, small angle
neutron scattering, and specific heat experiments.

The existence of different length scales for the two sets of
bands, rather directly demonstrated for the coherence length (for
example by scanning tunnelling spectroscopy and analysis of field
dependent specific heat data), but not yet for the penetration
depth (naturally, there is only one local field distribution for
the whole of MgB$_2$), strongly suggests that any phenomenological
model should be extended to a two band version before being
applied to the analysis of data obtained on MgB$_2$ $-$ otherwise
obtained ``best fit'' single band values of basic length scales
will be effective values that acquire a field dependence due to
the field dependence of the relative contributions of the two sets
of bands, although the bands specific length scales likely are not
field dependent. In the case of the London model, which was used
so successfully for the analysis of torque, small angle neutron
scattering, muon spin rotation, magnetization, and many other data
measured e.g.\ on cuprate superconductors, it is rather unclear
whether such an extension can be successful. This is so because of
the large overlap of the $\pi$ bands vortex cores throughout most
of the mixed state phase diagram, which violates a basic
assumption of the London model. It is clear, that while the
puzzling mixed state anisotropy effects may be understood on a
qualitative level, there are still open questions concerning their
more quantitative description.

From the study of cuprate superconductors is known that the mixed
state can be strongly affected by thermal fluctuations and
inevitably present material disorder, leading to phase diagrams
containing more transition lines than the upper and lower critical
fields. The material disorder leads for example to a destruction
of the (quasi)ordered vortex lattice in fields lower than the
upper critical field. This happens in an order-disorder
transition, which was also observed in conventional
superconductors, such as NbSe$_2$. This transition, manifested by
a peak in the critical current density with peculiar history
effects is observed in MgB$_2$ as well, by torque magnetometry, as
well as various other measurements. MgB$_2$ is interesting in
terms of these additional ``vortex matter'' transitions as well,
because of the intermediate importance of thermal fluctuations,
compared to cuprate and conventional superconductors. A melting
transition driven predominantly by thermal fluctuations has not
been observed clearly in MgB$_2$ so far though. We note that there
are possible influences of the two band superconductivity on the
non-equilibrium properties and these additional transitions of
vortex matter as well. However, a direct influence has not been
indicated by any experiment so far, and is thus left for future
studies.

We would like to conclude this chapter by briefly commenting on
the consequences of the results that were reviewed for
applications of MgB$_2$. The relatively high $T_{\mathrm{c}}$ and
critical current densities (not affected by ``weak links'' across
boundaries), coupled with the easy and cheap processing (as
compared to the high $T_{\mathrm{c}}$ cuprate superconductors) has
created also an intense technical interest in MgB$_2$
\cite{Larbalestier01rev}. However, the relatively large electronic
anisotropy of the $\sigma$ bands, which inevitable becomes the
relevant anisotropy when the magnetic field is increased to values
depressing the superconductivity in the isotropic $\pi$ bands,
rather degrades the possibility of granular MgB$_2$ material to
carry high critical current densities $j_{\mathrm{c}}$ in high
magnetic fields: correspondingly, $j_{\mathrm{c}}$ has been found
to be a rapidly decreasing function of the applied magnetic field
\cite{Buzea01}. In high quality single crystals, magnetization
curves are almost reversible in a wide region even at low
temperature [see Fig.\ \ref{PE1}a)]. It will be important to
maximize the pinning properties by increasing the disordered
vortex phase, which is restricted to a small region below
$H_{\mathrm{c2}}$ in clean material (see Fig.\ \ref{phd775deg}),
through the controlled introduction of uncorrelated material
defects. This increases the processing requirements for MgB$_2$
based materials for high field applications. However, the
potential of strongly affecting also the anisotropy by selectively
tuning the intraband impurity scattering rates
\cite{Gurevich03,Ribeiro03} may lead to applicability enhancements
that cannot yet be predicted.

\section*{Acknowledgements}
\label{ack}

We would like to thank J.~Jun, J.~Karpinski, and S.~M. Kazakov of
ETH Z{\"{u}}rich, D.\ Di Castro, H.~Keller, S.~Kohout, and J.~Roos
of the University of Z{\"{u}}rich, A.~Wisniewski of the Polish
Academy of Sciences, and P.~Miranovi{\'{c}} of Okayama University
for their help and comments during our collaborations at the time
when the torque experiments being reviewed here were performed. We
are grateful to D.~Larbalestier and H.~J. Choi for the permission
to use figures from their papers \cite{Larbalestier01rev,Choi02b}
(Figs.\ \ref{struc} and \ref{FS} in this chapter). MA would like
to thank B.~Batlogg and H.~R. Ott for valuable comments he
received during the preparation of his thesis \cite{AngstPhD}, in
which a large part of the torque results reviewed here was already
presented, and furthermore F.~Bouquet, A.~Golubov, V.~G. Kogan,
and P.~Miranovi{\'{c}} for fruitful discussions of various details
of the general picture presented in this chapter and for the
communication of calculated numerical data and formulas.

%\bibliographystyle{prsty}
%\bibliography{MgB2,da_hist,da_pin,da_th_em,da_th_mi,da_mate,own,Y124}

\begin{thebibliography}{100}

\bibitem{Russell53}
V. Russell, R. Hirst, F.~A. Kanda, and A.~J. King, Acta Cryst.
{\bf 6},  870
  (1953).

\bibitem{Nagamatsu01}
J. Nagamatsu, N. Nakagawa, T. Muranaka, Y. Zenitani, and J.
Akimitsu, Nature
  {\bf 410},  63  (2001).

\bibitem{Buzea01}
C. Buzea and T. Yamashita, Supercond. Sci. Technol. {\bf 14},
R115  (2001).

\bibitem{note_phononTc}
The belief that the transition temperature $T_{\text{c}}$ of
conventional
  phonon mediated superconductors was limited to some $30$ to $40\,{\text{K}}$
  was widespread. See for example {\em High-Temperature Superconductivity},
  edited by V.~L. Ginzburg and D.~A. Kirzhnits (Consultants Bureau, New York,
  1982); R.~J. Cava, J. Am. Ceram. Soc. {\bf 83}, 5 (2000). It should be
  mentioned, however, that in papers trying to theoretically calculate
  $T_{\text{c}}$, it was generally cautioned that any such numerical limits can
  only be obtained {\em for a given class of materials} [see, e.g., W.~L.
  McMillan, Phys. Rev. {\bf 167}, 331 (1967)], while new classes of materials
  may have much higher $T_{\text{c}}$. A relatively old idea concerning such a
  different material with much higher $T_{\text{c}}$ was for example the
  speculative metallic hydrogen [see, e.g., N.~W. Ashcroft, Phys. Rev. Lett.
  {\bf 21}, 1748 (1968)].

\bibitem{Budko01}
S.~L. Bud'ko, G. Lapertot, C. Petrovic, C.~E. Cunningham, N.
Anderson, and
  P.~C. Canfield, Phys. Rev. Lett. {\bf 86},  1877  (2001).

\bibitem{Kotegawa01}
H. Kotegawa, K. Ishida, Y. Kitaoka, T. Muranaka, and J. Akimitsu,
Phys. Rev.
  Lett. {\bf 87},  127001  (2001).

\bibitem{DiCastro03}
D.~Di Castro, R.~Khasanov, D.~Eshchenko, A.~Shengelaya, I.~M.
Savic, K.~Conder,
  S.~M. Kazakov, J.~Karpinski, M.~Angst, J.~Roos, and H.~Keller, in
  preparation.

\bibitem{note_Keller03}
See, e.g., H.~Keller, Physica B {\bf 326}, 283 (2003), and
references therein.

\bibitem{Schmidt01}
H. Schmidt, J.~F. Zasadzinski, K.~E. Gray, and D.~G. Hinks, Phys.
Rev. B {\bf
  63},  220504(R)  (2001).

\bibitem{Wang01c}
Y. Wang, T. Plackowski, and A. Junod, Physica C {\bf 355},  179
(2001).

\bibitem{Bouquet01b}
F. Bouquet, R.~A. Fisher, N.~E. Phillips, D.~G. Hinks, and J.~D.
Jorgensen,
  Phys. Rev. Lett. {\bf 87},  047001  (2001).

\bibitem{Hinks01}
D.~G. Hinks, H. Claus, and J.~D. Jorgensen, Nature {\bf 411},  457
(2001).

\bibitem{Liu01}
A.~Y. Liu, I.~I. Mazin, and J. Kortus, Phys. Rev. Lett. {\bf 87},
087005
  (2001).

\bibitem{Choi02}
H.~J. Choi, D. Roundy, H. Sun, M.~L. Cohen, and S.~G. Louie, Phys.
Rev. B {\bf
  66},  020513(R)  (2002).

\bibitem{Choi02b}
H.~J. Choi, D. Roundy, H. Sun, M.~L. Cohen, and S.~G. Louie,
Nature {\bf 418},
  758  (2002).

\bibitem{Canfield03}
P.~C. Canfield and G.~W. Crabtree, Physics Today {\bf 56},  34
(2003).

\bibitem{AngstPhD}
M. Angst, PhD thesis, ETH Z{\"u}rich, Switzerland, 2003. Available online at \\
  {http://e-collection.ethbib.ethz.ch/show?type=diss\&nr=14887}.

\bibitem{Larbalestier01rev}
D. Larbalestier, A. Gurevich, D.~M. Feldmann, and A. Polyanskii,
Nature {\bf
  414},  368  (2001).

\bibitem{Suhl59}
H. Suhl, B.~T. Matthias, and L.~R. Walker, Phys. Rev. Lett. {\bf
3},  552
  (1959).

\bibitem{Carlson70}
J.~R. Carlson and C.~B. Satterthwaite, Phys. Rev. Lett. {\bf 24},
461  (1970).

\bibitem{Butler76}
W.~H. Butler and P.~B. Allen,  in {\em Superconductivity in D- and
F- Metals},
  edited by D.~H. Douglass (Plenum, New York, 1976).

\bibitem{Binnig80}
G. Binnig, A. Baratoff, H.~E. Hoenig, and J.~G. Bednorz, Phys.
Rev. Lett. {\bf
  45},  1352  (1980).

\bibitem{Kresin90}
V.~Z. Kresin and S.~A. Wolf, Phys. Rev. B {\bf 41},  4278  (1990).

\bibitem{Pickett89}
W.~E. Pickett, Rev. Mod. Phys. {\bf 61},  433  (1989).

\bibitem{Belashchenko01}
K.~D. Belashchenko, M. van Schilfgaarde, and V.~P. Antropov, Phys.
Rev. B {\bf
  64},  092503  (2001).

\bibitem{An01}
J.~M. An and W.~E. Pickett, Phys. Rev. Lett. {\bf 86},  4366
(2001).

\bibitem{Kortus01}
J. Kortus, I.~I. Mazin, K.~D. Belashchenko, V.~P. Antropov, and
L.~L. Boyer,
  Phys. Rev. Lett. {\bf 86},  4656  (2001).

\bibitem{Bohnen01}
K.-P. Bohnen, R. Heid, and B. Renker, Phys. Rev. Lett. {\bf 86},
5771  (2001).

\bibitem{Kong01}
Y. Kong, O.~V. Dolgov, O. Jepsen, and O.~K. Andersen, Phys. Rev. B
{\bf 64},
  020501(R)  (2001).

\bibitem{Yildirim01}
T. Yildirim, O. G{\"u}lseren, J.~W. Lynn, C.~M. Brown, T.~J.
Udovic, Q. Huang,
  N. Rogado, K.~A. Regan, M.~A. Hayward, J.~S. Slusky, T. He, M.~K. Haas, P.
  Khalifah, K. Inumaru, and R.~J. Cava, Phys. Rev. Lett. {\bf 87},  037001
  (2001).

\bibitem{note_gainedel}
Its three natural electrons, plus the one it gained from
magnesium.

\bibitem{Boeri02}
L. Boeri, G.~B. Bachelet, E. Cappelluti, and L. Pietronero, Phys.
Rev. B {\bf
  65},  214501  (2002).

\bibitem{Shukla03}
A. Shukla, M. Calandra, M. {d'Astuto}, M. Lazzeri, F. Mauri, C.
Bellin, M.
  Krisch, J. Karpinski, S.~M. Kazakov, J. Jun, D. Daghero, and K. Parlinski,
  Phys. Rev. Lett. {\bf 90},  095506  (2003).

\bibitem{note_nonadiab}
Recently, it was argued, that nonadiabatic effects due to
closeness of the
  Fermi level to the top of the $\sigma$ bands are important for the pairing in
  MgB$_2$ and that nonadiabatic pairing is the primary source of the high
  $T_{\text{c}}$. See E.~Cappelluti, S.~Ciuchi, C.~Grimaldi, L.~Pietronero, and
  S.~Str{\"a}ssler, Phys. Rev. Lett. {\bf 88}, 117003 (2002), and Ref.\
  \cite{Boeri02}. However, the absence of a boron isotope effect on the
  $\sigma$ bands effective mass \cite{DiCastro03} points against the importance
  of nonadiabatic effects for the superconductivity in MgB$_2$.

\bibitem{Mazin03}
I.~I. Mazin and V.~P. Antropov, Physica C {\bf 385},  49  (2003).

\bibitem{note_anisTc}
It is quite well known for a long time that anisotropy in the
pairing
  interaction raises $T_{\text{c}}$: The notion that pairing anisotropy is
  beneficial to high $T_{\text{c}}$ appeared the first time, as an almost
  incidental remark, in Ref.\ \cite{Markowitz63}. A quite general theoretical
  study of interaction anisotropy and $T_{\text{c}}$ can be found, for example,
  in O.~T. Valls and M.~T. B{\'e}al-Monod, Phys. Rev. B {\bf 51}, 8438 (1995).

\bibitem{Golubov97}
A.~A. Golubov and I.~I. Mazin, Phys. Rev. B {\bf 55},  15146
(1997).

\bibitem{Mazin02}
I.~I. Mazin, O.~K. Andersen, O. Jepsen, O.~V. Dolgov, J. Kortus,
A.~A. Golubov,
  A.~B. {Kuz'menko}, and D. {van der~Marel}, Phys. Rev. Lett. {\bf 89},  107002
   (2002).

\bibitem{Yelland02}
E.~A. Yelland, J.~R. Cooper, A. Carrington, N.~E. Hussey, P.~J.
Meeson, S. Lee,
  A. Yamamoto, and S. Tajima, Phys. Rev. Lett. {\bf 88},  217002  (2002).

\bibitem{Carrington03}
A. Carrington, P.~J. Meeson, J.~R. Cooper, L. Balicas, N.~E.
Hussey, E.~A.
  Yelland, S. Lee, A. Yamamoto, S. Tajima, S.~M. Kazakov, and J. Karpinski,
  submitted to Phys.\ Rev.\ Lett., preprint on cond-mat/0304435.

\bibitem{Uchiyama02}
H. Uchiyama, K.~M. Shen, S. Lee, A. Damascelli, D.~H. Lu, D.~L.
Feng, Z.-X.
  Shen, and S. Tajima, Phys. Rev. Lett. {\bf 88},  157002  (2002).

\bibitem{Szabo01}
P. Szab{\'o}, P. Samuely, J. Kamar{\'{\i}}k, T. Klein, J. Marcus,
D. Fruchart,
  S. Miraglia, C. Marcenat, and A.~G.~M. Jansen, Phys. Rev. Lett. {\bf 87},
  137005  (2001).

\bibitem{Laube01}
F. Laube, G. Goll, J. Hagel, H. v.~Lohneysen, D. Ernst, and T.
Wolf, Europhys.
  Lett. {\bf 56},  296  (2001).

\bibitem{Gonnelli02}
R.~S. Gonnelli, D. Daghero, G.~A. Ummarino, V.~A. Stepanov, J.
Jun, S.~M.
  Kazakov, and J. Karpinski, Phys. Rev. Lett. {\bf 89},  247004  (2002).

\bibitem{Giubileo01}
F. Giubileo, D. Roditchev, W. Sacks, R. Lamy, and J. Klein,
Europhys. Lett.
  {\bf 58},  764  (2002).

\bibitem{Giubileo01a}
F. Giubileo, D. Roditchev, W. Sacks, R. Lamy, D. Thanh, J. Klein,
S. Miraglia,
  D. Fruchart, J. Marcus, and P. Monod, Phys. Rev. Lett. {\bf 87},  177008
  (2001).

\bibitem{Iavarone02}
M. Iavarone, G. Karapetrov, A.~E. Koshelev, W.~K. Kwok, G.~W.
Crabtree, D.~G.
  Hinks, W.~N. Kang, E.-M. Choi, H.~J. Kim, H.-J. Kim, and S.~I. Lee, Phys.
  Rev. Lett. {\bf 89},  187002  (2002).

\bibitem{Eskildsen02}
M.~R. Eskildsen, M. Kugler, S. Tanaka, J. Jun, S.~M. Kazakov, J.
Karpinski, and
  {\O}. Fischer, Phys. Rev. Lett. {\bf 89},  187003  (2002).

\bibitem{Tsuda01}
S. Tsuda, T. Yokoya, T. Kiss, Y. Takano, K. Togano, H. Kito, H.
Ihara, and S.
  Shin, Phys. Rev. Lett. {\bf 87},  177006  (2001).

\bibitem{Schmidt02}
H. Schmidt, J.~F. Zasadzinski, K.~E. Gray, and D.~G. Hinks, Phys.
Rev. Lett.
  {\bf 88},  127002  (2002).

\bibitem{Chen01}
X. Chen, M. Konstantinovic, J. Irwin, D. Lawrie, and J. Franck,
Phys. Rev.
  Lett. {\bf 87},  157002  (2001).

\bibitem{Quilty02}
J.~W. Quilty, S. Lee, A. Yamamoto, and S. Tajima, Phys. Rev. Lett.
{\bf 88},
  087001  (2002).

\bibitem{Souma03}
S. Souma, Y. Machida, T. Sato, T. Takahashi, H. Matsui, S.-C.
Wang, H. Ding, A.
  Kaminski, J.~C. Campuzano, S. Sasaki, and K. Kadowaki, Nature {\bf 423},  65
  (2003).

\bibitem{Bouquet01c}
F. Bouquet, Y. Wang, R.~A. Fisher, D.~G. Hinks, J.~D. Jorgensen,
A. Junod, and
  N.~E. Phillips, Europhys. Lett. {\bf 56},  856  (2001).

\bibitem{Sologubenko02b}
A.~V. Sologubenko, J. Jun, S.~M. Kazakov, J. Karpinski, and H.~R.
Ott, Phys.
  Rev. B {\bf 66},  014504  (2002).

\bibitem{Golubov02}
A.~A. Golubov, J. Kortus, O.~V. Dolgov, O. Jepsen, Y. Kong, O.~K.
Andersen,
  B.~J. Gibson, K. Ahn, and R.~K. Kremer, J. Phys.: Cond. Mat. {\bf 14},  1353
  (2002).

\bibitem{Simon01}
F. Simon, A. J{\'a}nossy, T. Feh{\'e}r, F. Mur{\'a}nyi, S. Garaj,
L. Forr{\'o},
  C. Petrovic, S. L.{ Bud'ko}, G. Lapertot, V.~G. Kogan, and P.~C. Canfield,
  Phys. Rev. Lett. {\bf 87},  047002  (2001).

\bibitem{Budko01b}
S.~L. Bud'ko, V.~G. Kogan, and P.~C. Canfield, Phys. Rev. B {\bf
64},
  180506(R)  (2001).

\bibitem{Delima01}
O.~F. {de~Lima}, R.~A. Ribeiro, M.~A. Avila, C.~A. Cardoso, and
A.~A. Coelho,
  Phys. Rev. Lett. {\bf 86},  5974  (2001).

\bibitem{Lima01a}
O.~F. de~Lima, C.~A. Cardoso, R.~A. Ribeiro, M.~A. Avila, and
A.~A. Coelho,
  Phys. Rev. B {\bf 64},  144517  (2001).

\bibitem{Papavassiliou02}
G. Papavassiliou, M. Pissas, M. Fardis, M. Karayanni, and C.
Christides, Phys.
  Rev. B {\bf 65},  012510  (2002).

\bibitem{Kim02}
K.~H.~P. Kim, J.-H. Choi, C.~U. Jung, P. Chowdhury, H.-S. Lee,
M.-S. Park,
  H.-J. Kim, J.~Y. Kim, Z. Du, E.-M. Choi, M.-S. Kim, W.~N. Kang, S.-I. Lee,
  G.~Y. Sung, and J.~Y. Lee, Phys. Rev. B {\bf 65},  100510(R)  (2002).

\bibitem{Xu01}
M. Xu, H. Kitazawa, Y. Takano, J. Ye, K. Nishida, H. Abe, A.
Matsushita, and G.
  Kido, Appl. Phys. Lett. {\bf 79},  2779  (2001).

\bibitem{Lee01}
S. Lee, H. Mori, T. Masui, Y. Eltsev, A. Yamamoto, and S. Tajima,
J. Phys. Soc.
  Jpn. {\bf 70},  2255  (2001).

\bibitem{Pradhan01}
A.~K. Pradhan, Z.~X. Shi, M. Tokunaga, T. Tamegai, Y. Takano, K.
Togano, H.
  Kito, and H. Ihara, Phys. Rev. B {\bf 64},  212509  (2001).

\bibitem{Tinkham_intro}
M. Tinkham, {\em Introduction to Superconductivity} (McGraw Hill,
New York,
  1996).

\bibitem{note_effmass}
In fact, in the microscopic derivations of the anisotropic
Ginzburg-Landau
  theory, the effective mass tensor is given by $(m^{-1})_{i,j} = 1/(2
  E_{\text{F}})\, \langle \varphi ^2 (\overrightarrow{k})\, v_{\text{F}}^i
  (\overrightarrow{k}) v_{\text{F}}^i (\overrightarrow{k}) \rangle$, where
  $\langle \ldots \rangle$ denotes an average over the Fermi surface, and
  $\varphi$ describes the $\overrightarrow{k}-$dependence of the
  superconducting gap. Therefore, the effective mass tensor components actually
  entering the GL free energy functional are not identical to the band
  effective masses. See Ref.\ \cite{Gorkov64}.

\bibitem{Ginzburg_Landau50}
V.~L. Ginzburg and L.~D. Landau, Zh. Eksp. Teor. Fiz. {\bf 20},
1064  (1950).

\bibitem{Ginzburg52}
V.~L. Ginzburg, Zh. Eksp. Teor. Fiz. {\bf 23},  236  (1952).

\bibitem{Caroli63}
C. Caroli, P.~G. {De~Gennes}, and J. Matricon, Phys. Konden.
Mater. {\bf 1},
  176  (1963).

\bibitem{Gorkov64}
L.~P. {Gor'kov} and T.~K. Melik-Barkhudarov, Zh. Eksperim. i Teor.
Fiz. {\bf
  45},  1493  (1963), [English transl.: Soviet Phys. JETP {\bf 18}, 1031
  (1964)].

\bibitem{Tilley65}
D. R. Tilley, Proc. Phys. Soc. London {\bf 86}, 678 (1965); {\bf
86}, 678
  (1965).

\bibitem{Blatter92}
G. Blatter, V.~B. Geshkenbein, and A.~I. Larkin, Phys. Rev. Lett.
{\bf 68},
  875  (1992).

\bibitem{Muto77}
Y. Muto, K. Noto, H. Nakatsuji, and N. Toyota, Nuovo Cimento B
{\bf 38},  503
  (1977).

\bibitem{Takanaka75}
K. Takanaka, Phys. Stat. Sol. B {\bf 68},  623  (1975).

\bibitem{Teichler75}
H. Teichler, Phys. Stat. Sol. B {\bf 72},  211  (1975).

\bibitem{MgB2anisPRL02}
M. Angst, R. Puzniak, A. Wisniewski, J. Jun, S.~M. Kazakov, J.
Karpinski, J.
  Roos, and H. Keller, Phys. Rev. Lett. {\bf 88},  167004  (2002).

\bibitem{Karpinski03SST}
J. Karpinski, M. Angst, J. Jun, S.~M. Kazakov, R. Puzniak, A.
Wisniewski, J.
  Roos, H. Keller, A. Perucchi, L. Degiorgi, M.~R. Eskildsen, P. Bordet, L.
  Vinnikov, and A. Mironov, Supercond. Sci. Technol. {\bf 16},  221  (2003).

\bibitem{Angst03MgB2PhyC}
M. Angst, R. Puzniak, A. Wisniewski, J. Roos, H. Keller, P.
Miranovi{\'{c}}, J.
  Jun, S.~M. Kazakov, and J. Karpinski, Physica C {\bf 385},  143  (2003).

\bibitem{Angst03Rio}
M. Angst, D. {Di~Castro}, R. Puzniak, A. Wisniewski, J. Jun, S.~M.
Kazakov, J.
  Karpinski, S. Kohout, and H. Keller, submitted to Physica C, prepreint on
  cond-mat/0304400.

\bibitem{Sologubenko02}
A.~V. Sologubenko, J. Jun, S.~M. Kazakov, J. Karpinski, and H.~R.
Ott, Phys.
  Rev. B {\bf 65},  180505(R)  (2002).

\bibitem{Budko02}
S.~L. {Bud'ko} and P.~C. Canfield, Phys. Rev. B {\bf 65},  212501
(2002).

\bibitem{Eltsev02}
Y. Eltsev, S. Lee, K. Nakao, N. Chikumoto, S. Tajima, N.
Koshizuka, and M.
  Murakami, Phys. Rev. B {\bf 65},  140501(R)  (2002).

\bibitem{Ferdeghini02}
C. Ferdeghini, V. Ferrando, V. Braccini, M.~R. Cimberle, D.
Marr{\'e}, P.
  Manfrinetti, A. Palenzona, and M. Putti, Eur. Phys. J. B {\bf 30},  147
  (2002).

\bibitem{Zehetmayer02}
M. Zehetmayer, M. Eisterer, J. Jun, S.~M. Kazakov, J. Karpinski,
A. Wisniewski,
  and H.~W. Weber, Phys. Rev. B {\bf 66},  052505  (2002).

\bibitem{Lyard02}
L. Lyard, P. Samuely, P. Szabo, T. Klein, C. Marcenat, L. Paulius,
K.~H.~P.
  Kim, C.~U. Jung, H.-S. Lee, B. Kang, S. Choi, S.-I. Lee, J. Marcus, S.
  Blanchard, A.~G.~M. Jansen, U. Welp, G. Karapetrov, and W.~K. Kwok, Phys.
  Rev. B {\bf 66},  180502(R)  (2002).

\bibitem{Welp03}
U. Welp, G. Karapetrov, W.~K. Kwok, G.~W. Crabtree, C. Marcenat,
L. Paulius, T.
  Klein, J. Marcus, K.~H.~P. Kim, C.~U. Jung, H.-S. Lee, B. Kang, and S.-I.
  Lee, Phys. Rev. B {\bf 67},  012505  (2003).

\bibitem{Machida03}
Y. Machida, S. Sasaki, H. Fujii, M. Furuyama, I. Kakeya, and K.
Kadowaki, Phys.
  Rev. B {\bf 67},  094507  (2003).

\bibitem{Posazhennikova02}
A.~I. Posazhennikova, T. Dahm, and K. Maki, Europhys. Lett. {\bf
60},  134
  (2002).

\bibitem{Miranovic03}
P. Miranovi{\'{c}}, K. Machida, and V.~G. Kogan, J. Phys. Soc.
Jpn. {\bf 72},
  221  (2003).

\bibitem{Gurevich03}
A. Gurevich, Phys. Rev. B in print (1 May 2003 issue), preprint on
  cond-mat/0212129.

\bibitem{Dahm02}
T. Dahm and N. Schopohl, cond-mat/0212188.

\bibitem{Golubov03}
A.~A. Golubov and A.~E. Koshelev, submitted to Phys.\ Rev.\ B,
preprint on
  cond-mat/0303237.

\bibitem{note_ABC}
Four MgB$_2$ single crystals were used in all torque studies,
labeled A, B, C,
  and D, throughout this chapter. Crystal A with the smallest volume of about
  $4\times 10^{-4}\, {\text{mm}}^3$ was measured in the home made torque
  magnetometer with a maximum field of $14\,{\text{kOe}}$ and shaking option.
  Crystals B and C were measured with the Quantum Design PPMS with torque
  option, operating with fields up to $90\,{\text{kOe}}$. Crystal D was
  measured in the same magnetometer as crystal A, but using a capacitive torque
  sensor [C.~Rossel, M.~Willemin, A.~Gasser, H.~Bothuizen, G.~I. Meijer, and
  H.~Keller, Rev. Sci. Instrum. {\bf 69}, 3199 (1998)] instead of a
  piezoresistive one.

\bibitem{Karpinski03phyc}
J. Karpinski, S.~M. Kazakov, J. Jun, M. Angst, R. Puzniak, A.
Wisniewski, and
  P. Bordet, Physica C {\bf 385},  42  (2003).

\bibitem{MgB2PE}
M. Angst, R. Puzniak, A. Wisniewski, J. Jun, S.~M. Kazakov, and J.
Karpinski,
  Phys. Rev. B {\bf 67},  012502  (2003).

\bibitem{Willemin98b}
M. Willemin, C. Rossel, J. Brugger, M.~H. Despont, H. Rothuizen,
P. Vettiger,
  J. Hofer, and H. Keller, J. Appl. Phys. {\bf 83},  1163  (1998).

\bibitem{Willemin98}
M. Willemin, C. Rossel, J. Hofer, H. Keller, A. Erb, and E.
Walker, Phys. Rev.
  B {\bf 58},  R5940  (1998).

\bibitem{Willemin98a}
M. Willemin, A. Schilling, H. Keller, C. Rossel, J. Hofer, U.
Welp, W.~K. Kwok,
  R.~J. Olsson, and G.~W. Crabtree, Phys. Rev. Lett. {\bf 81},  4236  (1998).

\bibitem{note_incompatible}
Form and angular regime of the deviation from a straight line are
incompatible
  with a superconducting signal, because the maximum of a superconducting
  signal would be much closer to $90\,{\text{deg}}$ for an anisotropic
  superconductor.

\bibitem{note_AHLTfixedT}
This is what justifies our AGLT based analysis {\em a posteriori}.
Note that
  closer to $T_{\mathrm{c}}$, there are deviations of the angular dependence of
  $H_{\mathrm{c2}}$ from AGLT predictions [Eq.\ (\ref{Hc2_theta})]. This is
  discussed later.

\bibitem{note_Gi}
Note that there are at least two inequivalent definitions of the
Ginzburg
  number used in the literature. Throughout this thesis, we use the definition
  also used in Ref.\ \cite{Mikitik01}.

\bibitem{Forgan02}
E.~M. Forgan, S.~J. Levett, P.~G. Kealey, R. Cubitt, C.~D.
Dewhurst, and D.
  Fort, Phys. Rev. Lett. {\bf 88},  167003  (2002).

\bibitem{Eremenko02}
V.~V. Eremenko, V.~A. Sirenko, Y.~A. Shabakayeva, R. Schleser, and
P.~L.
  Gammel, Low Temp. Phys. {\bf 28},  6  (2002).

\bibitem{Mikitik01}
G.~P. Mikitik and E.~H. Brandt, Phys. Rev. B {\bf 64},  184514
(2001).

\bibitem{Schneiderbook}
T. Schneider and J.~M. Singer, {\em Phase Transition Approach to
High
  Temperature Superconductivity} (Imperial College Press, London, 2000).

\bibitem{Lee72}
P.~A. Lee and S.~R. Shenoy, Phys. Rev. Lett. {\bf 28},  1025
(1972).

\bibitem{Ullah90}
S. Ullah and A.~T. Dorsey, Phys. Rev. Lett. {\bf 65},  2066
(1990).

\bibitem{Welp91}
U. Welp, S. Fleshler, W.~K. Kwok, R.~A. Klemm, V.~M. Vinokur, J.
Downey, B.
  Veal, and G.~W. Crabtree, Phys. Rev. Lett. {\bf 67},  3180  (1991).

\bibitem{Wilkin93}
N.~K. Wilkin and M.~A. Moore, Phys. Rev. B {\bf 48},  3464
(1993).

\bibitem{Lawrie94}
I.~D. Lawrie, Phys. Rev. B {\bf 50},  9456  (1994).

\bibitem{Dulcic03}
A. Dulcic, M. Pozek, D. Paar, E.-M. Choi, H.-J. Kim, W.~N. Kang,
and S.-I. Lee,
  Phys. Rev. B {\bf 67},  020507(R)  (2003).

\bibitem{Buzdin92}
A. Buzdin and D. Feinberg, Physica C {\bf 220},  74  (1992).

\bibitem{note_scalingangle}
Since the function
$\mathrm{F}(\mathrm{A}(T-T_{\mathrm{c2}})/(TH)^{2/3})$ is
  not fixed for different angles of anisotropic superconductors and depends
  slightly on the angle [J.~M. Calero, J.~C. Granada, and E.~Z. da Silva, Phys.
  Rev. B {\bf 56} 6114 (1997)] in order to get very accurate experimental data
  of $H{\mathrm{c2}}(\theta)$, reliable e.g.\ for the discussion of tiny
  details of the deviation of $H{\mathrm{c2}}(\theta)$ from the anisotropic
  Ginzburg-Landau angular dependence \cite{Angst03Rio}, the angular dependence
  of $\mathrm{F}(\mathrm{A}(T-T_{\mathrm{c2}})/(TH)^{2/3})$ should be taken
  into account in the determination of $H_{\mathrm{c2}}(\theta)$.

\bibitem{note_anisC}
The upper critical field dataset of crystal C is not complete, at
  $34\,{\text{K}}$, the datasets of the two crystals also agree, within error
  bars.

\bibitem{Miranovic_pc}
P. Miranovi{\'{c}}, private communication.

\bibitem{note_resistLyard}
The determination of $H_{\mathrm{c2}}$ from electrical transport
measurements
  is problematic in the case of MgB$_2$. In Ref.\ \cite{Lyard02}, a criterion
  was chosen that gave $H_{\mathrm{c2}}$ values matching those determined by
  specific heat and ac susceptibility measured on the same crystal (for $H\|c$
  and $H\|ab$). This criterion is different from the onset criterion chosen in
  earlier studies.

\bibitem{note2}
We note that the parameter values for which best agreement with
Ref.\
  \cite{Golubov03} is seen are $4$-$5\%$ higher than those of the AGLT ``best
  fit''.

\bibitem{Helfand66}
E. Helfand and N.~R. Werthamer, Phys. Rev. {\bf 147},  288
(1966).

\bibitem{Cooper_pc}
J.~R. Cooper, private communication.

\bibitem{Usadel70}
K.~D. Usadel, Phys. Rev. Lett. {\bf 25},  507  (1970).

\bibitem{Ribeiro03}
R.~A. Ribeiro, S.~L. {Bud'ko}, C. Petrovic, and P.~C. Canfield,
Physica C {\bf
  384},  227  (2003).

\bibitem{Perucchi02}
A. Perucchi, L. Degiorgi, J. Jun, M. Angst, and J. Karpinski,
Phys. Rev. Lett.
  {\bf 89},  097001  (2002).

\bibitem{Koshelev03}
A.~E. Koshelev and A.~A. Golubov, Phys. Rev. Lett. {\bf 90},
177002  (2003).

\bibitem{Bouquet02}
F. Bouquet, Y. Wang, I. Sheikin, T. Plackowski, A. Junod, S. Lee,
and S.
  Tajima, Phys. Rev. Lett. {\bf 89},  257001  (2002).

\bibitem{Kogan02}
V.~G. Kogan, Phys. Rev. B {\bf 66},  020509  (2002).

\bibitem{Golubov02b}
A.~A. Golubov, A. Brinkman, O.~V. Dolgov, J. Kortus, and O.
Jepsen, Phys. Rev.
  B {\bf 66},  054524  (2002).

\bibitem{Brinkman02}
A. Brinkman, A.~A. Golubov, and H. Rogalla, Phys. Rev. B {\bf 65},
180517(R)
  (2002).

\bibitem{Manzano02}
F. Manzano and A. Carrington, Phys. Rev. Lett. {\bf 88},  047002
(2002).

\bibitem{Perkins02}
G.~K. Perkins, J. Moore, Y. Bugoslavsky, L.~F. Cohen, J. Jun,
S.~M. Kazakov, J.
  Karpinski, and A.~D. Caplin, Supercond. Sci. Technol. {\bf 15},  1156
  (2002).

\bibitem{Zeldov94}
E. Zeldov, A.~I. Larkin, M. Konczykowski, B. Khaykovich, D. Majer,
V.~B.
  Geshkenbein, and V.~M. Vinokur, Physica C {\bf 235-240},  2761  (1994).

\bibitem{Farrell88}
D.~E. Farrell, C.~M. Williams, S.~A. Wolf, N.~P. Bansal, and V.~G.
Kogan, Phys.
  Rev. Lett. {\bf 61},  2805  (1988).

\bibitem{Farrell89}
D.~E. Farrell, S. Bonham, J. Foster, Y.~C. Chang, P.~Z. Jiang,
K.~G.
  Vandervoort, D.~J. Lam, and V.~G. Kogan, Phys. Rev. Lett. {\bf 63},  782
  (1989).

\bibitem{Zech96}
D. Zech, C. Rossel, L. Lesne, H. Keller, S.~L. Lee, and J.
Karpinski, Phys.
  Rev. B {\bf 54},  12535  (1996).

\bibitem{Hofer98}
J. Hofer, J. Karpinski, M. Willemin, G.~I. Meijer, E.~M. Kopnin,
R. Molinski,
  H. Schwer, C. Rossel, and H. Keller, Physica C {\bf 297},  103  (1998).

\bibitem{Kogan81}
V.~G. Kogan, Phys. Rev. B {\bf 24},  1572  (1981).

\bibitem{Kogan88}
V.~G. Kogan, M.~M. Fang, and S. Mitra, Phys. Rev. B {\bf 38},
11958  (1988).

\bibitem{Kogan88b}
V.~G. Kogan, Phys. Rev. B {\bf 38},  7049  (1988).

\bibitem{Hao_Clem91}
Z. Hao and J.~R. Clem, Phys. Rev. Lett. {\bf 67},  2371  (1991).

\bibitem{Hao91b}
Z. Hao and J.~R. Clem, Phys. Rev. B {\bf 43},  7622  (1991).

\bibitem{CommentTakahashi}
M. Angst, R. Puzniak, A. Wisniewski, J. Roos, H. Keller, and J.
Karpinski,
  submitted to Phys. Rev. B, preprint on cond-mat/0206407.

\bibitem{note_Brandtshake}
E.~H. Brandt and G.~P. Mikitik, Phys. Rev. Lett. {\bf 89}, 027002
(2002); G.~P.
  Mikitik and E.~H. Brandt, Phys. Rev. B {\bf 67}, 104511 (2003).

\bibitem{note_smallirr}
Only in the case of the lowest temperatures and fields measured
there was a
  small irreversibility remaining even after shaking, always limited to angles
  in the interval from $85$ to $95\,{\text{deg}}$. In these cases, the data
  within the interval from $85$ to $95\,{\text{deg}}$ were excluded from the
  analysis with Eq.\ (\ref{tau_rev}).

\bibitem{Takahashi02}
K. Takahashi, T. Atsumi, N. Yamamoto, M. Xu, H. Kitazawa, and T.
Ishida, Phys.
  Rev. B {\bf 66},  012501  (2002).

\bibitem{note1}
For $H < H_{c2}$, $\gamma _H$ is generalized to mean the
anisotropy of the
  coherence lengths $\xi$.

\bibitem{Kogan02b}
V.~G. Kogan, Phys. Rev. Lett. {\bf 89},  237005  (2002).

\bibitem{note_epsilon}
Note that these $\epsilon_{\lambda , H}(\theta)$ correspond to the
  $\Theta_{\lambda , H}(\theta)$ in Ref.\ \cite{Kogan02b}.

\bibitem{Nakai02}
N. Nakai, M. Ichioka, and K. Machida, J. Phys. Soc. Jpn. {\bf 71},
23  (2002).

\bibitem{Yethiraj93}
M. Yethiraj, H.~A. Mook, G.~D. Wignall, R. Cubitt, E.~M. Forgan,
S.~L. Lee,
  D.~M. Paul, and T. Armstrong, Phys. Rev. Lett. {\bf 71},  3019  (1993).

\bibitem{Campbell88}
L.~J. Campbell, M.~M. Doria, and V.~G. Kogan, Phys. Rev. B {\bf
38},  2439
  (1988).

\bibitem{Cubitt03}
R. Cubitt, S. Levett, S.~L. {Bud'ko}, N.~E. Anderson, and P.~C.
Canfield, Phys.
  Rev. Lett. {\bf 90},  157002  (2003).

\bibitem{Eskildsen_pc}
M. R. Eskildsen, private communication.

\bibitem{Babaev02}
E. Babaev, Phys. Rev. Lett. {\bf 89},  067001  (2002).

\bibitem{Gurevich03b}
A. Gurevich and V.~M. Vinokur, Phys. Rev. Lett. {\bf 90},  047004
(2003).

\bibitem{Schilling96}
A. Schilling, R.~A. Fisher, N.~E. Phillips, U. Welp, D. Dasgupta,
W.~K. Kwok,
  and G.~W. Crabtree, Nature {\bf 382},  791  (1996).

\bibitem{Giamarchi94}
T. Giamarchi and P. {Le~Doussal}, Phys. Rev. Lett. {\bf 72},  1530
(1994).

\bibitem{Khaykovich96}
B. Khaykovich, E. Zeldov, D. Majer, T.~W. Li, P.~H. Kes, and M.
Konczykowski,
  Phys. Rev. Lett. {\bf 76},  2555  (1996).

\bibitem{Giller97}
D. Giller, A. Shaulov, R. Prozorov, Y. Abulafia, Y. Wolfus, L.
Burlachkov, Y.
  Yeshurun, E. Zeldov, V.~M. Vinokur, J.~L. Peng, and R.~L. Greene, Phys. Rev.
  Lett. {\bf 79},  2542  (1997).

\bibitem{note_Avraham}
We follow convention and treat melting and order-disorder
transition
  separately, although there is evidence that they form one line of phase
  transitions, which just changes character from thermally driven to disorder
  driven. See N.~Avraham {\em{et al.}}, Nature {\bf 411}, 451 (2001).

\bibitem{Ravikumar01}
G. Ravikumar, V.~C. Sahni, A.~K. Grover, S. Ramakrishnan, P.~L.
Gammel, D.~J.
  Bishop, E. Bucher, M.~J. Higgins, and S. Bhattacharya, Phys. Rev. B {\bf 63},
   024505  (2001).

\bibitem{Marchevsky01}
M. Marchevsky, M.~J. Higgins, and S. Bhattacharya, Nature {\bf
409},  591
  (2001).

\bibitem{Ling01}
X.~S. Ling, S.~R. Park, B.~A. McClain, S.~M. Choi, D.~C. Dender,
and J.~W.
  Lynn, Phys. Rev. Lett. {\bf 86},  712  (2001).

\bibitem{note_Sr124}
See e.g. M.~Angst, S.~M. Kazakov, J.~Karpinski, A.~Wisniewski,
R.~Puzniak, and
  M.~Baran , Phys. Rev. B {\bf 65}, 094518 (2002), and references therein.

\bibitem{Roy00}
S.~B. Roy, P. Chaddah, and S. Chaudhary, Phys. Rev. B {\bf 62},
9191  (2000).

\bibitem{Pissas02}
M. Pissas, S. Lee, A. Yamamoto, and S. Tajima, Phys. Rev. Lett.
{\bf 89},
  097002  (2002).

\bibitem{note_Hirr_crit}
A criterion $\tau_{\text{irr}}/H\sin(2\theta)=10^{-6}
  {\text{dyn}}\,{\text{cm}}\,{\text{Oe}}^{-1}$ was used for defining
  $H_{\text{irr}}$.

\bibitem{Bean62}
C.~P. Bean, Phys. Rev. Lett. {\bf 8},  250  (1962).

\bibitem{Puzni_prep}
R.~Puzniak {\em et al.}, unpublished.

\bibitem{note_Banerjee99}
See, e.g., S.~S. Banerjee, N.~G. Patil, S. Ramakrishnan, A.~K.
Grover, S.
  Bhattacharya, P.~K. Mishra, G. Ravikumar, T.~V.~C. Rao, V.~C. Sahni, M.~J.
  Higgins, C.~V. Tomy, G. Balakrishnan, and D.~M. Paul, Phys. Rev. B {\bf 59},
  6043 (1999).

\bibitem{Markowitz63}
D. Markowitz and L.~P. Kadanoff, Phys. Rev. {\bf 131},  563
(1963).

\end{thebibliography}

\newcommand{\noopsort}[1]{} \newcommand{\printfirst}[2]{#1}
  \newcommand{\singleletter}[1]{#1} \newcommand{\switchargs}[2]{#2#1}

\end{document}